\documentclass[aps,floatfix,a4paper,showpacs,twocolumn,nofootinbib,superscriptaddress,10pt]{revtex4}
\usepackage{graphicx,float}\usepackage{graphicx,float}
\usepackage[all]{xy}
\usepackage{amsmath}
\usepackage{amssymb}
\usepackage{color}
\usepackage{epsfig}		
\usepackage{graphicx,epstopdf}
\usepackage{subfigure}
\usepackage{pdfpages}
\usepackage[colorlinks,hyperindex]{hyperref}

\hypersetup{
colorlinks,
citecolor=blue,
linkcolor=black,
urlcolor=black,
}

\newcommand{\bes}{\begin{subequations}}
\newcommand{\ees}{\end{subequations}}
\def\ben{\begin{eqnarray}}
\def\een{\end{eqnarray}}
\def\be{\begin{equation}}
\def\ee{\end{equation}}

\begin{document}

\title{Entropic information of dynamical AdS/QCD holographic models}
\author{Alex E. Bernardini} 
\affiliation{Departamento de F\'isica, Universidade Federal de S\~ao Carlos,
PO Box 676, 13565-905, S\~ao Carlos, SP, Brasil}
\email{alexeb@ufscar.br}
\author{Rold\~ao da Rocha}
\affiliation{Centro de Matem\'atica, Computa\c c\~ao e Cogni\c c\~ao, Universidade Federal do ABC, UFABC, 09210-580, Santo Andr\'e, Brazil}
\email{roldao.rocha@ufabc.edu.br}

\begin{abstract}
The Shannon based conditional entropy that underlies five-dimensional Einstein-Hilbert gravity coupled to a dilaton field is investigated in the context of dynamical holographic AdS/QCD models.
Considering the UV and IR dominance limits of such AdS/QCD models, the conditional entropy is shown to shed some light onto the meson classification schemes, which corroborate with the existence of light-flavour mesons of lower spins in Nature. 
Our analysis is supported by a correspondence between statistical mechanics and information entropy which establishes the physical grounds to the Shannon information entropy, also in the context of statistical mechanics, and provides some specificities for accurately extending the entropic discussion to continuous modes of physical systems.
From entropic informational grounds, the conditional entropy allows one to identify the lower experimental/phenomenological occurrence of higher spin mesons in Nature.
Moreover, it introduces a quantitative theoretical apparatus for studying the instability of high spin light-flavour mesons.
\end{abstract}
\maketitle


\section{Introduction}

The AdS/CFT correspondence, relating weakly coupled SUGRA in 
 anti-de Sitter space (AdS$_5$) to the $\mathcal{N} = 4$ super Yang-Mills theory in
the boundary, plays a prominent role in the strong coupling regime \cite{maldacena,maldacena1998}. In fact, strongly coupled gauge theories can be described by higher-dimensional theories of gravity. 
In the low energy and non-perturbative regime
of QCD, usually named as AdS/QCD \cite{hooft}, a weakly coupled AdS$_5$ background -- dual to conformal symmetry of the 4D field theory -- is assumed.
Indeed, the asymptotic freedom makes QCD to be approximately
conformal at high energies. Thereafter, bulk fields, which are dual to QCD operators through the AdS/CFT correspondence, are included into the 5D framework. AdS/QCD has a reasonable agreement with experimental results for both ratios of meson masses and structure constants, whenever the bulk setup 
regards fundamental features of low energy QCD \cite{Karch:2006pv,Colangelo:2008us,Chelabi:2015gpc}. 

Gauge/gravity duality has been applied to study aspects of QCD, as to compare the existing predictions to QCD
data \cite{PDG} and to study theories that are also based on those data \cite{Csaki,Kinar}. 
In this context, the ultraviolet (UV) and infrared (IR) limits in AdS/QCD can be relevant. In particular,
the UV limit reproduces the scaling behavior of QCD scattering amplitudes,
whereas the IR regime in the extra dimension, at the QCD scale provided by $\Lambda _{%
\text{QCD}}$, is able to carry out the discrete hadron spectra as well as the mass gap. 
The so-called hard-wall QCD predicts squared masses of particles that are quadratic functions of principal and spin quantum numbers 
\cite{hw,hwph,hwph1}. Nevertheless, it does not comply with the linear Regge behavior of the mass mesonic spectra \cite{ani00}, that can be derived by considering an AdS$_{5}$ bulk geometry 
through the addition of an IR suppressed bulk scalar field \cite{gremm,De_Wolfe_PRD_2000} -- the dilaton background -- that depends on the extra dimension, that behaves as an energy scale, and that realises conformal symmetry breaking, accordingly \cite{Karch:2006pv}. The boundary vacuum properties, as for instance the existence of condensates and quark confinement, are dynamically comprised by solutions of the Einstein field equations coupled to a dilaton field, without the need of any additional background field. This is the underlying setup of soft-wall models, that 
indeed generate linear Regge trajectories \cite{erlich,Karch:2006pv,dePaula:2012jh,dePaula:2010yu,dePaula:2009za,dePaula:2008fp}, whose phenomenology was also explored in Ref. \cite{Galow:2009kw}.

In this context, once proposed to realize chiral symmetry breaking in the vacuum and asymptotic chiral symmetry restoration in high excitation states, holographic QCD models have been successful in describing Regge trajectories for vector and axial-vector meson fields.
In a type II superstring theory with the spectra of vector and axial-vector mesons, the dependence of the Regge trajectories parameters on the metric parameters in correspondence to a realistic holographic QCD model has been identified \cite{SongHe2010}. Otherwise, even being relevant in providing an explanation for spontaneously chiral symmetry breaking, the string theory Sakai-Sugimoto model \cite{Sakai2005}, fails to generate the linear Regge behavior. Still more pragmatic, the so-called {\em bottom-up} approach has been relevant in generating the linear Regge behavior for meson and baryon spectra \cite{Karch:2006pv,Andreev2006,Kruczenski,dePaula:2009za,dePaula:2008fp}.
 
In AdS/QCD models, a conformal bulk metric has conformal symmetry that can be broken, thus corresponding to a dual conformal field theory. Moreover, the AdS/QCD correspondence is introduced when a cut-off, or hard wall, is considered at some point in the bulk. Hence, the gravity branch of the theory is defined only from this cut-off, ensuring confinement, since, in fact, the scale cut-off corresponds to a mass gap. Confinement is implemented
either through a hard cutoff in the extra dimension, or by a dynamical
cutoff, such as a nontrivial dilaton background \cite{Csaki}. It consists into the soft wall model that accurately derives the linear spectrum, in both radial and spin
directions \cite{Karch:2006pv}. 

These models shall be here explored, in the context of the lattice approach of the Shannon information entropy, realized by the conditional entropy \cite{glst,glsow}. 
Moreover, the correct continuum limit of the information entropy shall be derived, in agreement with a ground correspondence between statistical mechanics and information theory, also identified along recently considered \emph{ad hoc} approaches \cite{roldao,plb2016}. 
Within this framework, the formation of light-flavour mesons in AdS/QCD soft wall models, in a procedure relating their spins to their respective conditional entropy, can be scrutinized. 

The concept of informational entropy regards the unpredictability of information in a system and has been applied 
to compact systems \cite{glst,glsow}. The lattice approach of Shannon information entropy measures the shape complexity of spatially localized physical configurations, providing an informational measure of the complexity of
a system that has spatial profile, in terms of its frequency (through its Fourier transform) \cite{glst,glsow,glgra}.
The lower the information entropy, the less information encoded in the frequency modes, is required to represent 
the shape. The lattice approach of Shannon information entropy has been successfully applied in several contexts, encompassing non-equilibrium dynamics \cite{glst}, 
 stability bounds for compact objects \cite{glsow,Gleiser:2015rwa} and cosmology \cite{glgra}. The conditional entropy encrypts an extension of the Shannon entropy to a continuum limit of the system modes of 
spatially-localized density functions, based upon their
Fourier transforms, that can provide, for example, reliable bounds on the
stability of physical systems. 

It shall be shown that the statistical mechanics content \cite{glst, glsow} of the modal fraction is the normalized structure factor, up to a constant. It provides a robust physical basis for what one knows 
as configurational entropy, establishing its limitations in the continuum limit, which is provided by the solid grounds of statistical mechanics. 
The conditional entropy satisfies all the required properties of entropy and will be here employed in the soft wall model of AdS/QCD to demonstrate that the lower the spin, the higher the informational organization is, in the configuration structure of light flavour mesonic systems. 

This work is organised as follows: in Sect. II, the statistical mechanics grounds that underlie the information entropy are  provided, and the range of their applications in the continuum limit of system frequency modes is discussed. The well-known collective coordinates 
and the structure factor in statistical mechanics, associated to the energy density of the system, shall be shown to be the main ingredients
to establish the relationship between the conditional entropy with the thermodynamical entropy. The well-known modal fraction, in the information entropy setup, is established as the normalized  structure factor, constructed upon the collective coordinates in statistical mechanics, and associated to the energy density of the system, as the seminal constituent that endows the conditional entropy with thermodynamical grounds. 
 In order to classify excited light-flavour mesons in the context of the information entropy, in Sect.~III, the 
framework for dynamical AdS/QCD holographic models is introduced, deriving the collective coordinates and the structure factors for arbitrary meson spins. Hence the information entropy of the system is analysed by means of the conditional entropy. From the point of view of the conditional entropy, light-flavour mesons with lower spins are then shown to be more stable than higher spin mesonic states. The Sakai-Sugimoto model is further employed and compared with the our previous results. Conclusively, the conditional entropy is shown to point towards the more frequent occurrence of lower spin light-flavour mesons in Nature. 
Our final conclusions are drawn in Sect. IV, also analysing the  higher spin mesons contributions, due to large $N_c$ suppression, in the conditional entropy setup.
 
\section{Conditional entropy and Shannon information entropy: statistical mechanics grounds}
 
The entropic information is realised by the conditional entropy, 
for spatially-localised solutions in field theories \cite{glst}. It has been latterly 
propounded as a precise paradigm based upon the Shannon information entropy. The so-called configurational entropy (CE) has a prominent importance in the lattice approach, containing finite frequency modes, and it has been 
applied with success in a variety of problems \cite{glst,glsow,glgra}. 
The entropic information paradigm realised by the CE shall be here applied to study the entropic information content and the stability of light-flavour mesonic field configurations, within the AdS/QCD framework that shall be established in Sect. \ref{ILI}. The entropic information measure is useful for quantifying the informational and organizational content encoded by the energy density of physical solutions \cite{glst}. In this context, besides the least action principle, physical systems do not only minimise the corresponding action, being the entropic information also optimised. Hence, entropic-informationally stable physical systems always have an optimised configurational organization, accordingly \cite%
{glst}. States of higher CE either demand more energy to be produced, or are more rarely observed than their configurationally stable counterparts, or both. 

The CE is based upon the Shannon information paradigm \cite{glst}, being originally 
defined, for a discrete system consisting of $n$ modes, by $S_c = -{{\sum_{k=1}^n}} f_k\,\ln(f_k)$, where $f_n$ denotes some probability density function that describes the studied system. Historically, $S_c$ is called the information entropy \cite{shannon}. The CE, indeed, encrypts a well-established setup, wherein the entropy is defined as (minus) the sum ${{\sum_{k=1}^n}} f_k\,{\rm log}_b(f_k)$, and encompasses additional entropy formulations, namely the Nat entropy and the Hartley entropy, respectively for $b=e$ and $b=10$, besides the Shannon entropy for $b = 2$.

Such definition is compatible with the interpretation that CE defines the best lossless compression of any exchange of information, namely, any communication inside, or even out of, the system. The higher the CE, the higher the unpredictability of information content in the system is. Hence, the CE carries the entropic information in physical system 
configurations, and can be used, further, to bound any physical parameter in the system, as well as to corroborate with experimental, phenomenological or observational data. It is worth to mention that if the system conveys $n$ modes with equal weights, then the relation $f_k = 1/n$ holds for the relative probability mass functions. Entropy is maximum when the symbols have equal probabilities. Hence, 
the discrete CE presents a maximum value at $S_c = \ln n$. On the other hand, when the system presents a single mode, then $S_c = 0$, as it is expected \cite%
{glst}. 

The expression for $S_c$ indicates the system non-determination degree, from an informational
viewpoint, and is analogous to the Boltzmann-Gibbs entropy in
thermodynamics, expressing the disorder degree in a physical
system. Moreover, Kolmogorov established the relationship between the conditional entropy with the thermodynamical entropy in physics \cite{kolmo}.
This setup provides a reliable measure of bounds on the
stability and on the entropic information content of physical systems. In fact, since the continuum limit of Shannon entropy 
lacks some required properties that underlie any useful definition of entropy, the links between 
the conditional entropy and its statistical mechanics grounds must be clarified. The lattice approach of Shannon information entropy is an entropy
of shape, namely, an informational measure of the complexity of
a system that has spatial profile, in terms of its frequency \cite{glst,glsow,glgra}.
The lower the information entropy, the less information, encoded in the frequency modes, is required to represent 
the shape \cite{glst,glsow}. The continuum limit of the information entropy has been taken as an \emph{ad hoc} continuum limit \cite{roldao,plb2016}, however this limit is defined as being the differential entropy, that lacks 
fundamental properties \cite{shannon,kolmo,Martin}. For example, the configurational entropy -- or differential entropy -- is not a limit of the Shannon entropy for 
the number $n\to\infty$ of modes, since it differs from the limit of the Shannon entropy. The complementary part is what is called information dimension, describing the growth rate of the Shannon entropy in successively finer discretizations of the space. 
Hence, care should be taken when applying the differential entropy to physical systems.
In order to circumvent those limitations regarding the continuum limit, the lattice approach of the conditional entropy can be considered, as it has been successfully applied in Refs. \cite{glst,glsow,glgra}.

A close relationship between the information entropy and statistical thermodynamics 
can be derived. The most general formula for the thermodynamic entropy of a thermodynamic system is the Gibbs entropy,
${S=-k_{{B} }\sum_i p_{i}\ln p_{i}\,}$
where $k_B$ is the Boltzmann constant, and $p_i$ denotes the probability of a microstate. The von Neumann entropy is also defined, 
${S=-k_{{B} }\,{\rm {Tr}}(\rho \ln \rho )\,}$, where $\rho$ is the density matrix associated to the quantum mechanical system.
It is worth to point out that, in thermodynamics, the entropy takes into account macroscopic measurements, that play the role (but is not) of a probability distribution, which defines the information entropy. Nevertheless, thermodynamics can be linked to information theory by the Boltzmann equation, 
$ 
{S=k_{B}\ln(W)}$, 
where $W$ is the number of microstates that can yield the given macrostate. The probability of a given microstate is assumed to be $p_i = 1/W$, namely, all microstates are equally likely. Hence, the information entropy of a system is a measure of lacking information that is necessary to determine a microstate. In this way, thermodynamic entropy in statistical mechanics is, thus, proportional\footnote{The constant of proportionality is the Boltzmann constant.} to the quantity of Shannon information that is requisite to define the microscopic state of a given system, that stays 
uncommunicated by a description only in terms of the macroscopic variables of classical thermodynamics \cite{Martin}. A robust formal 
setup, with possibilities involving entanglement entropy is provided in Ref. \cite{queiroz}.

Although the Shannon entropy is strictly restricted to random variables that take discrete values, its continuum limit can be regarded, by considering a continuous random variable with probability density function. The conditional entropy is an extension of the Shannon discrete entropy. Nevertheless, a subtlety must be considered, since the continuum limit have some limitations in what concern that properties of entropy. 
The Fourier transform of the energy density\footnote{The following definition is quite general and holds for high-dimensions \cite{glsow}. However, one opportunely adopts the spatial coordinate $z$, in particular, to denote the bulk coordinate, that shall be used further in AdS/QCD models.}, 
\begin{equation}
\varrho(\omega)=\frac{1}{\sqrt{2\pi }}\int \;e^{i\omega z}\rho(z)\,dz,
\label{collectivecoordinates}
\end{equation}
is employed for constructing the analogue continuum limit of the collective coordinates in statistical mechanics, that are defined by $\rho(z)\sim\sum e^{-i\omega_n z}\varrho(\omega_n)$, namely, the discretized version of Eq. (\ref{collectivecoordinates}) for periodic functions.  Here $n$ here denotes the number of modes in the physical system. 
The choice of the energy density is indeed suitable for defining the conditional entropy, since it is spatially localised and encodes the physics of the model. In statistical mechanics, the energy density, in the space of frequencies, has modes $\omega_n$ that compose the system, being the normalised Fourier transform of the energy density operator. Hence, the normalized structure factor is constructed upon the collective coordinates, reading, 
\begin{equation}
f(\omega_n)=\frac{\langle\;\left\vert \varrho(\omega_n)\right\vert ^{2}\;\rangle}{\;s_n}, \label{collective1}
\end{equation} where  
\begin{equation}
\label{sfactor}
s_n \sim \sum_{j=1}^n\langle\;\left\vert \varrho(\omega_n)\right\vert ^{2} \rangle\,.
\end{equation} 
 The structure factor measures the relative weight carried by each mode $\omega$, driving the physical system. The structure factor is here defined, regarding the finite amount of modes in that context of statistical mechanics. In the limit when $n \to \infty$, and taking into account Eq. (\ref{collectivecoordinates}), the correlation reads  
 \begin{equation}
\label{sfactor}
s=\lim_{n\to\infty}s_n={\displaystyle{\int_{-\infty}^\infty}}
 \langle\;\left\vert \varrho(\omega)\right\vert ^{2}\rangle d\omega\,,
 \end{equation} and Eq. (\ref{collective}) yields 
 \begin{equation}
 f(\omega)=\lim_{n\to\infty}f(\omega_n)=\frac{\langle\;\left\vert \varrho(\omega)\right\vert ^{2}\rangle}{s}\,. \label{collective}
 \end{equation}

Besides the Shannon's information theory, the continuum collective coordinate $\varrho(\omega)$ in Eq. (\ref{collectivecoordinates}) is an well-grounded 
 in statistical mechanics. In fact, it is the Fourier transform of the density operator $\rho(z)$. In a
homogeneous fluid, the average collective coordinate is trivial. However, even in a homogeneous system, the correlation of $f(\omega)$ is not trivial. The structure factor can be extended to the continuum limit, being defined as this correlation. Although the averaged density is constant in any homogeneous fluid, the energy density operator fluctuates from configuration to configuration. Hence, the structure factor in Eq. (\ref{sfactor}) is a measure of the fluctuations in the energy density. In fact, the structure factor is the Fourier transform of $\rho(z,z')\equiv \langle \rho(z)\rho(z')\rangle$, thus quantitatively measuring density fluctuations. Moreover, it is a measure of the tendency of the system towards homogenization. 

Hence, what is known in the literature as the modal fraction \cite{glst, glsow} is known in statistical mechanics as being the normalized structure factor ratio, up to a constant. The continuum limit of the information entropy has been taken as an \emph{ad hoc} continuum limit \cite{roldao,plb2016}, however this limit is defined as being the differential entropy, that lacks 
some required properties to have analogy with statistical mechanics. For example, the configurational entropy -- or differential entropy -- is not a limit of the Shannon entropy for 
the number $n\to\infty$ of modes, since it differs from the limit of the Shannon entropy. 
As abovementioned, in order to circumvent those limitations, the lattice approach \cite{glst,glsow,glgra} sets the conditional entropy symbolically defined by 
\begin{equation}
S_c[f]\sim -\int f(\omega)\,\ln\left(f(\omega)\right)\,d\omega, \label{conditional}
\end{equation} although this definition is solely valid for lattice approaches as the generalization of Shannon information entropy, for a number $n\;\slash\!\!\!\!\!\!\to\infty$ of modes \cite{glsow}. In fact, the complementary part is what is called information dimension, describing the growth rate of the Shannon entropy in successively finer discretizations of the space. 
Hence, care should be taken when applying the conditional entropy to physical systems.

In the next section, this setup shall be applied to derive the relationship between 
the conditional entropy and the spins of excited mesons in the AdS/QCD soft wall model. Furthermore, 
based upon the physical foundations of the Shannon entropy presented in this section, the measure of the spatial complexity
of the order parameter based on normalized structure factor ratio decomposition in Eq. (\ref{collectivecoordinates}) can be employed to identify the critical
points of the conditional entropy as a measure of stability.

\section{Conditional entropy of AdS/QCD models}
\label{ILI}

One starts by introducing the 5D action for the graviton-dilaton coupling in the Einstein frame:
\begin{equation}
\mathcal{S}=\int d^{5}x\sqrt{\left\vert g\right\vert }\left[ -\frac{R}{4}+%
\frac{1}{2}g_{MN}\nabla^M \phi\nabla^N \phi-V(\phi )\right] ,
\end{equation}
\noindent where $4\pi G=1$, $R$ denotes the scalar curvature, and the dilaton scalar field $%
\phi$ depends solely upon the extra dimension. In addition, $V(\phi )$ is a 
potential that describes the model, $g=\det (g_{MN})$, and one uses a conformal coordinate system with
\begin{equation}
g_{MN}=e^{-2A(z)}\eta _{MN}, \label{metric}
\end{equation}
\noindent a conformal metric, where the conformal coordinate is defined by $z = \int^y\exp\left(A(\tilde{y})\right)d\tilde{y}$, where $y$ denotes the extra dimension, $e^{2A}$ denotes the warp factor, $\eta_{MN}$ denotes the usual flat Lorentzian metric components, with $M, N=0,1,2,3,5$, and $\phi=\phi(z)$ generates a domain wall. 
The coordinate $z$ is interpreted as the holographic energy scale of the theory. 

The choice for the warp
factor 
\begin{equation}
A(z)=\ln z+C\left( z\right) \label{dobra}
\end{equation} is compatible with further requirements for the AdS/QCD model \cite{dePaula:2008fp}, 
where the function $C\left( z\right)$ describes non-conformal deformations
of the original AdS$_{5}$ bulk metric, and the (boundary) condition $C(0)=0$ restricts the bulk to asymptotically $\mathrm{AdS}_{5}$ geometries in the UV limit 
 \cite{Karch:2006pv,dePaula:2008fp}. 

By denoting $B^{\prime}(z)=dB(z)/dz$, for any quantity $B$ depending 
upon the variable $z$, Einstein equations ${G}_{MN}=%
{T}_{MN}$ can be employed together with the Euler-Lagrange ones $\nabla_M \phi\nabla^M\phi+V^{\prime }(\phi) =0$, yielding 
\begin{eqnarray}
3A^{\prime \prime }-3A^{\prime }{}^{2}-\frac{1}{2}\phi ^{\prime
2}-e^{-2A}V(\phi ) &=&0, 
\label{gremm1} \\
6A^{\prime }{}^{2}-\frac{1}{2}\phi ^{\prime 2}+e^{-2A}V(\phi ) &=&0, \label{gremm2} \\
\phi ^{\prime \prime }-3A^{\prime }\phi ^{\prime }-e^{-2A}\frac{dV}{d\phi }
&=&0. \label{fieldeqs}
\end{eqnarray}%
Hence, together with the effective 
Einstein equations, the equivalent expression 
\begin{equation}
\phi ^{\prime }=\sqrt{3(A^{\prime }{}^{2}+A^{\prime \prime})}~
\label{primeira}
\end{equation}%
 shows, in particular, that a constant dilaton solution
requires the warp factor to satisfy $A^{\prime \prime }=-A^{\prime }{}^{2}$, what is 
equivalent to assert that the boundary condition $C\left( 0\right) =0$ exact AdS$_{5}$ is the unique solution. Moreover, Eqs. (\ref{gremm1}) - (\ref{fieldeqs}) are equivalent to $$
\frac{d\phi(y)}{dy} = \frac{1}{2}\frac{dW(\phi)}{d\phi} \,\,\mbox{and}\,\, \frac{dA(y)}{dy} =\frac{1}{3}W(\phi),$$
for a superpotential $
W(\phi)$ that generates the potential as $V(\phi )=\frac{1}{8}\left( \frac{dW(\phi )}{d\phi }\right) ^{2}-\frac{1}{3}
W^2(\phi)$ \cite{De_Wolfe_PRD_2000}, when these quantities are written in terms of the extra-dimension $y$. Thus, the energy density $\rho$ reads 
\begin{equation}
\rho \sim e^{2A}\mathcal{L}, \label{eq.5}
\end{equation}
\noindent with $\mathcal{L}=\frac12g_{MN}\nabla^{M}\phi \nabla^{N}\phi -V(\phi )$, whereas the potential reads \cite{Kiritsis} 
\begin{equation}
V(\phi \left( z\right) )=\frac32e^{2A\left( z\right)}\left(A^{\prime
\prime }\left( z\right) -3A^{\prime }{}^{2}\left( z\right) \right)\,.
\label{potentialconformal}
\end{equation}%
Given a warp factor $A(z(y))$, Eqs. (\ref{primeira}) and
(\ref{potentialconformal}) immediately lead to solutions for the scalar field $\phi$ and the 
potential $V(\phi \left( z\right) )$.

In AdS/QCD models, a dynamical
cut-off, implemented by a dilaton background, yields confinement and the meson spectra
and decay constants can be, then, described. 
In order to study the CE and the entropic information content of the meson 
spectra for different spins in the dynamical AdS/QCD holographic model, suitable warp factors are usually chosen, in order to encompass the 
the usual IR ($z\rightarrow \infty$) and UV ($z\rightarrow 0$) regimes of AdS/QCD behavior of the metric (\ref{metric}), in a non-compact fifth dimension \cite{maldacena,maldacena1998}. The following warp factor \cite{Kinar,Kruczenski,tobias,Kiritsis} 
\begin{equation}
C_{a }(z)=z^{a }, \label{poly}
\end{equation} is a usual choice to be adopted into Eq. (\ref{dobra}), corresponding to an asymptotic AdS$_{5}$ bulk. 
 The range $a \geq 0$ [$a \geq 1$] is required for the (asymptotic) AdS$_{5}$ metric to govern the UV limit [for linear confinement]. Hence, Eq. (\ref{primeira}) yields the following bulk scalar field, with the matching condition $\phi (0)=0$ \cite{dePaula:2010yu,dePaula:2009za,dePaula:2008fp}: 
 \begin{widetext}
\begin{eqnarray}
&&\phi _{a }(z)=\frac{\sqrt{3}}{a }\left[ (1+a )\ln \left(
\frac{\left(a z^{{a}/{2}}+\sqrt{a +a ^{2}+a 
^{2}z^{a }}\right)e^{z^{a /2}\sqrt{a +a ^{2}+a ^{2}z^{a }}}}{\sqrt{a +a ^{2}}}\right)\right]\,.
\label{phi}
\end{eqnarray}\end{widetext}
Therefore, this solution makes the bulk scalar field to be a solution of (\ref{gremm1})-(\ref{fieldeqs}), and is led to a well-known solution \cite{Kiritsis}, for the particular case where $a =2$. The corresponding dilaton potential reads 
\begin{equation}
V\left( \phi _{a }\left( z\right) \right) =-\frac{3}{2}e^{2z^{a 
}}\left[ 4+7a z^{a }+a ^{2}z^{a }\left( 3z^{a 
}-1\right) \right] . \label{vphiz}
\end{equation}
The bulk scalar field has asymptotes
 $\lim_{z\rightarrow 0}\phi
_{a }(z)\sim z^{a /2}$ (UV) 
and $\lim_{z\rightarrow \infty}\phi _{a }(z)\sim z^{a }$ (IR). 
The mesonic excitation spectra, induced by the warp factor (\ref{poly}), was derived, e. g, in \cite{dePaula:2010yu,dePaula:2009za,dePaula:2008fp}. In addition, linear square mass
 Regge-like trajectories were studied in \cite{Karch:2006pv}, and shall be explored in the context of the CE.

The AdS$_5$ space induces a continuum of eigenmodes, with discrete spectrum, dictated by the 1-dimensional (the extra-dimension) quantum mechanical box analogue problem, with boundary conditions at a finite IR cut-off. In fact, a Kaluza-Klein-type splitting of higher spin $S>2$ string modes of the totally symmetric massive tensor fields 
$\psi_{N_1\ldots N_S}$ of rank $S$. Since these are the kinetic terms of higher spin fields in AdS$_5$, with gauge invariance
$\delta \psi_{N_1\ldots N_S}=\nabla_{({N_1}}\xi_{N_2\ldots N_S)}$ \cite{nastase}, the action reads $\int\! d^5x \!\sqrt{-g}\!\left[D^M\psi^{N_1\ldots N_S}D_M\psi_{N_1\ldots N_S}\!\!+\!\!M^2\psi_{N_1\ldots N_S}\psi^{N_1\ldots N_S}\right]$. Similarly, in the axial gauge in the dilaton-gravity background, it yields, as usual, the amplitudes $\psi _{n,S}$, which satisfy a 
 Schr\"odinger equation 
\begin{equation}\label{BBB1}\left[ -\partial _{z}^{2}+V_{QM}(z)\right] \psi_{n,S}=m_{n,S}^{2}\psi _{n,S},
\end{equation} where the analogue SUSY quantum mechanical potential reads 
\begin{equation}\label{BBBB}
V_{QM}(z)=\frac14B^{{\prime}^2}(z)-\frac12B^{\prime\prime }(z),\end{equation} with $B=\left( 2S-1\right) A+\phi$. This last condition is required as in the extended hard-wall model. The eigenvalues $m_{n,S}^{2}$ are the squared meson mass spectrum of the gauge theory in the boundary, namely, the domain wall itself \cite{Karch:2006pv,Csaki}. 
The Schr\"odinger equation yields, for $a =4$, the UV regime $\lim_{z\rightarrow 0}\phi(z)=
z^{2}$ \cite{erlich,Karch:2006pv,dePaula:2010yu,dePaula:2009za}. 

The linear Regge trajectories can be, therefore, studied for highly excited meson states. The leading IR regime yields the whole normalizable discrete spectrum with mass gap, for $a >1$ \cite{Kiritsis}. It is worth to emphasize that asymptotic values of the squared mass spectrum, corresponding to linear
trajectories, further require $a =2$ in the IR dominant component (\ref%
{poly}) of the warp factor, in order to generate a harmonic IR
potential. The trajectories yielded by Eq. (\ref{poly}) 
 have the mass spectra profile $m_{n}^{2}\propto \sigma n$, where $\sigma$ denotes the QCD string tension. For higher spin mesons, the semiclassical rotating relativistic QCD string yields $m_{S}^{2}\propto \sigma S$. Hence the gravity dual is modified in order that both regimes hold, namely, $m_{n}^{2}= m_{S}^{2}\equiv m_{n,S}^2$ \cite{Karch:2006pv,Kiritsis}. However, it can be further corrected, to encompass the existing QCD data \cite{PDG}, by adopting 
\begin{equation}
C(z)=\frac{\sqrt{3}+1}{2S+\sqrt{3}-1}\frac{z^2\Lambda _{\text{QCD}}^2}{%
e^{(1-z\Lambda _{\text{QCD}})}+1} \label{cosmologica}
\end{equation}%
as the warp factor, which shall be the model to be explored hereupon, in the context of the entropic information. 
{A {\em bottom-up} approach has already been proposed as to solve the gravity background with the above introduced wrap factor in a scenario that includes full back-reaction in the Einstein dilaton system \cite{SongHe2012}.
Supported by the Einstein equations of the graviton coupling with a real scalar dilaton field, a self-consistent framework is shown to the geometric background with black-hole for any given phenomenological holographic model.
In particular, the thermal QCD aspects have been considered to check how this warp factor works in realistic QCD.
Even if the confronted Regge behaviors of the hadron spectrum are slightly deformed \cite{SongHe2010}, those results ratify the relevance of the IR behavior as it is sensitive to low energy QCD, in particular, for the light meson spectrum \cite{SongHe2012}.
In addition, in the context of fitting the heavy quark potential confronted with several phenomenological models, and identifying the QCD beta function, a logarithmic correction term with an explicit IR cut-off in the deformed AdS$_5$ warp factor has been considered \cite{SongHe2011}.
In spite of not containing all the geometry information to support heavy and light quarks, the warp factor from (\ref{cosmologica}) has been demonstrated to work fine. 
In fact, a non-less realistic description of the meson mass spectrum with nearly universal Regge slopes has been achieved without any tuning of phenomenological parameters \cite{dePaula:2009za,dePaula:2008fp}.}

In fact, the meson spectrum arising from Eqs. (\ref{BBB1}) and (\ref{BBBB}) was derived in Ref. \cite{dePaula:2008fp}, in both the UV ($z\rightarrow 0$) and IR ($z\rightarrow \infty$) limits of the $V_{QM}(z)$ potential. The UV
regime, for $\lambda >1$, reads
\begin{equation}
V_{QM}(z)=z^{-2}\left({\rm a}_{0} +{\rm a}_{1} z^{\frac{%
\lambda }{2}}+{\rm a}_{2} z^{\lambda}+\cdots\right)  \label{vsmallz}
\end{equation}%
with coefficients 
\begin{eqnarray}
{\rm a}_{0}\left(S\right) &=&S^{2}-\frac14,\nonumber\\
{\rm a}_{1}\left(
S\right) &=&(3\lambda (\lambda +1))^{1/2}\left( S-\frac{\lambda}{4}\right)\,,\nonumber\\
{\rm a}_{2}\left( S\right) &=&\frac{\lambda}{4} \left[ 8S^{2}-4(\lambda +1)S+5\lambda +3\right].
\end{eqnarray} For $\lambda =4$, $\lim_{z\rightarrow 0}\phi(z)\propto z^{2}$ \cite{dePaula:2008fp}. 
On the other hand, linear trajectories for highly excited meson states, in the large-$N_{c}$ limit in the IR limit, yields
\begin{equation}
\lim_{z\rightarrow \infty }V_{QM}(z)=\frac{\lambda ^{2}}{4}(2S+\sqrt{3}-1)^{2}\;z^{2\lambda -2}.
\label{csm}
\end{equation}
It is worth to notice that the coefficient ot Eq. (\ref{csm}) is related to the coefficient in Eq. (\ref{cosmologica}). 
Furthermore, in the UV limit, the corresponding metric is the 
AdS$_{5}$ one. Nevertheless, for $z\gtrsim \Lambda _{\text{%
QCD}}^{-1}$ it regards the confining large-$z$ asymptotics of Eq. (\ref%
{poly}), with $a =2$. The associated dilaton field and
potential 
are then acquired, by numerically solving Eq. (\ref{primeira}). This determines the potential of the Schr\"odinger associated equation, and
the masses follow by solving the Schr\"odinger equation. In what follows the value $%
\Lambda _{\text{QCD}}=0.3$ GeV shall be adopted \cite{dePaula:2010yu,dePaula:2009za,dePaula:2008fp}. The running gauge coupling can be implemented into
Eq. (\ref{cosmologica}), for $z\ll \Lambda _{\text{QCD}}^{-1}$, inducing confinement at large $z$ \cite{dePaula:2008fp}.

In order to implement this setup into the dynamical AdS/QCD holographic model, and to define a further property related to mesons, for defining the CE in AdS/QCD models, the (warped) energy density must be taken into account. Thereafter, the continuum limit of the collective coordinates, regarding Eq. (\ref{collectivecoordinates}), 
 is defined as the warped collective coordinates
\begin{equation}
\varrho(\omega)=\frac{1}{\sqrt{2\pi }}\int \;e^{i\omega z+2A(z)}\rho(z)\,dz\,.
\end{equation}
\begin{widetext}
For the dynamical AdS/QCD holographic model compatible with linear Regge trajectories and QCD data, Eqs. (\ref{collectivecoordinates}) and (\ref{collective}) yield the profiles for the normalized structure factor ratio depicted in Figs. 1 and 2, respectively associated with 
the IR and UV limits of Eq. (\ref{cosmologica}).
\begin{figure}[H]
\begin{center}
\includegraphics[width=2.99in]{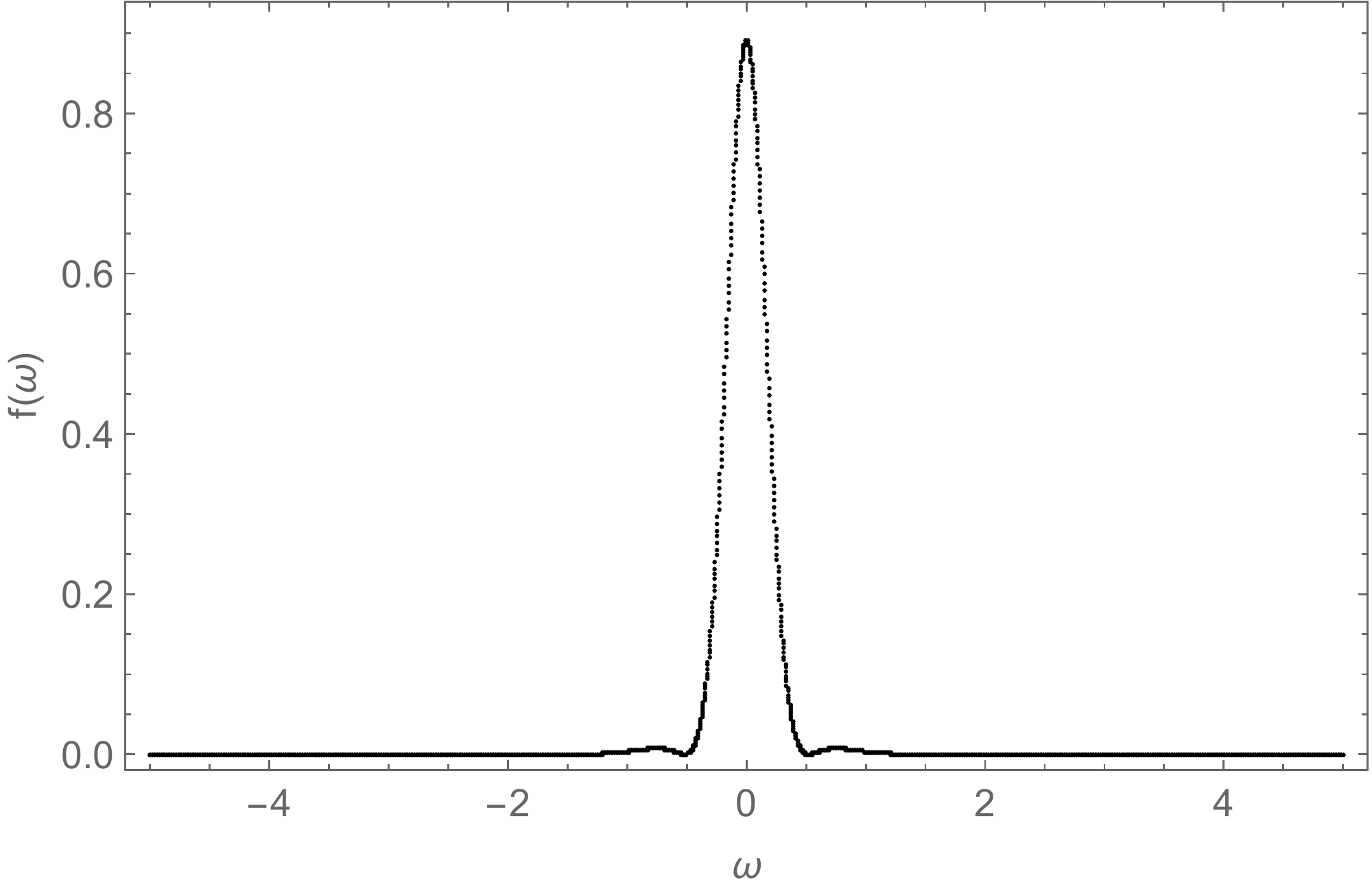}
\quad\quad
\includegraphics[width=2.99in]{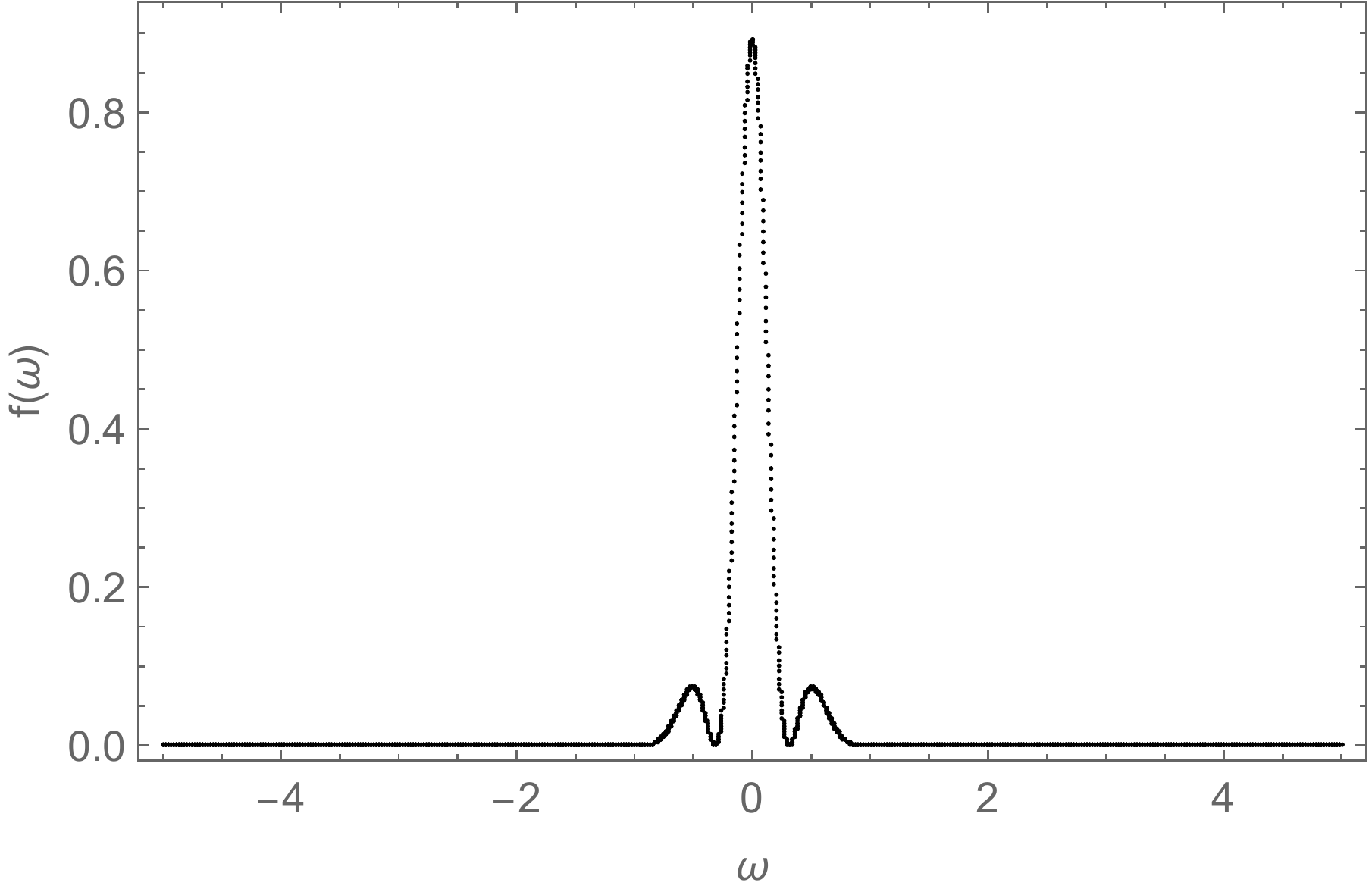}\\
\includegraphics[width=2.99in]{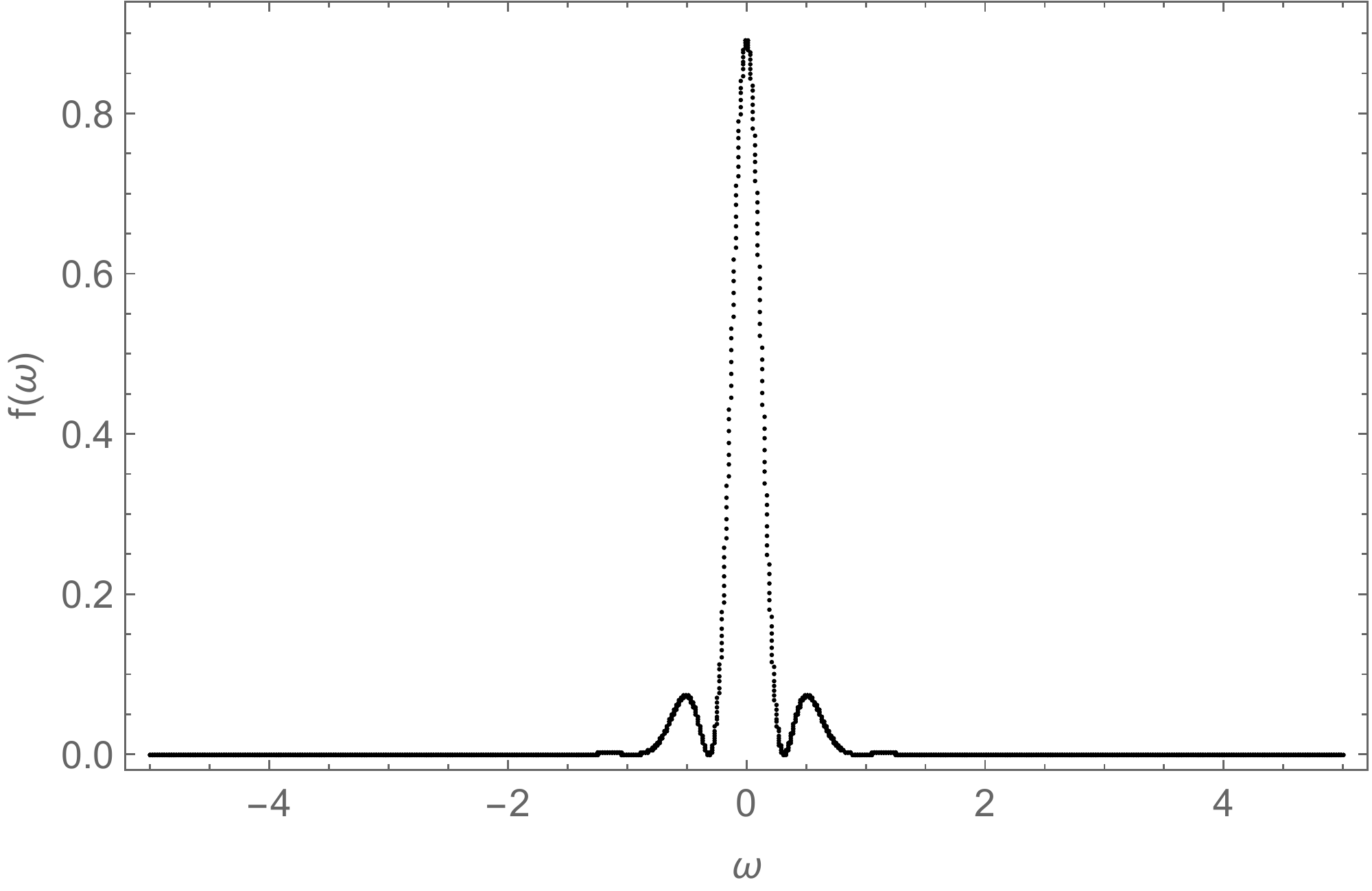}
\quad\quad
\includegraphics[width=2.99in]{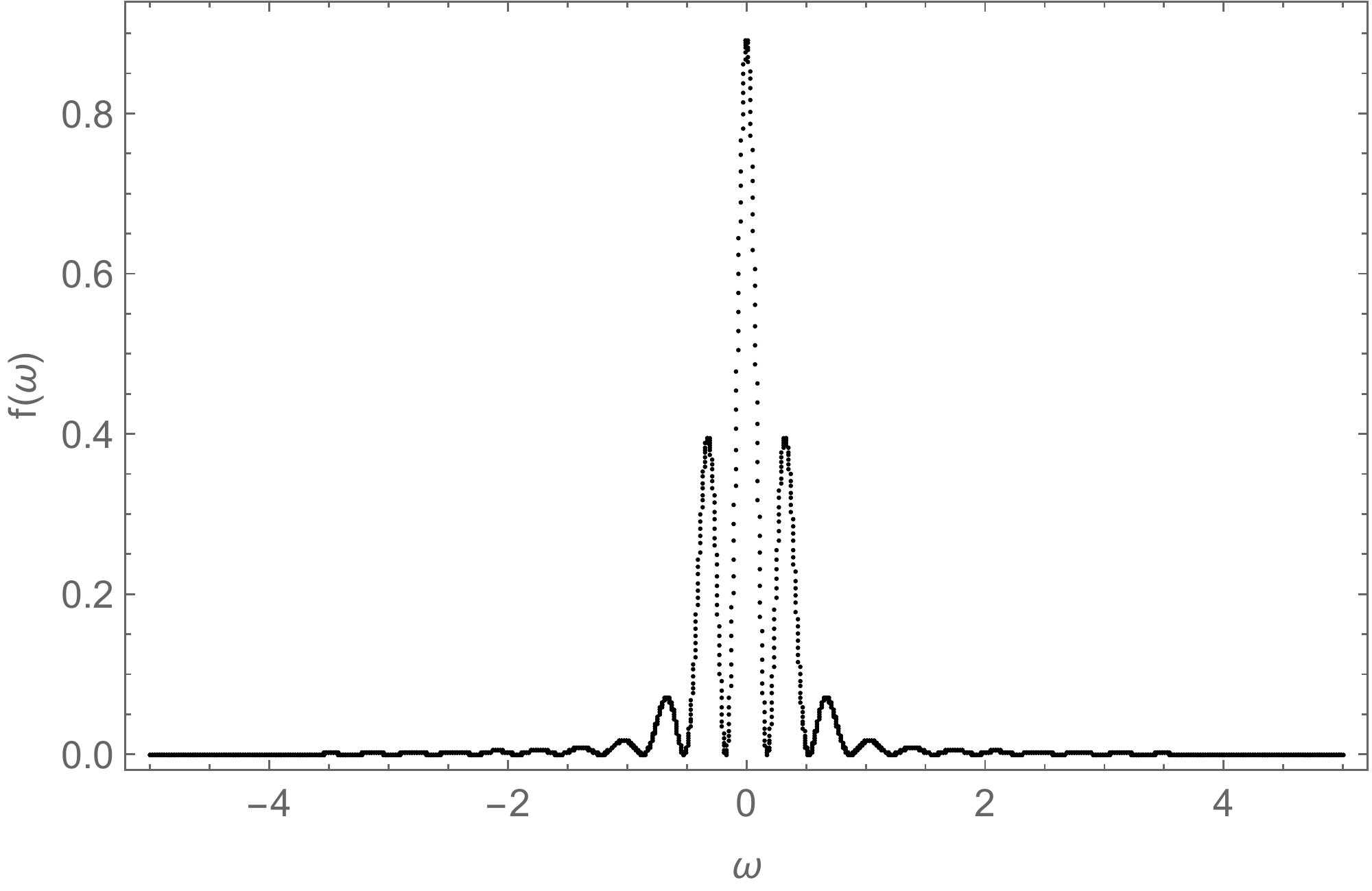}\\
\includegraphics[width=2.99in]{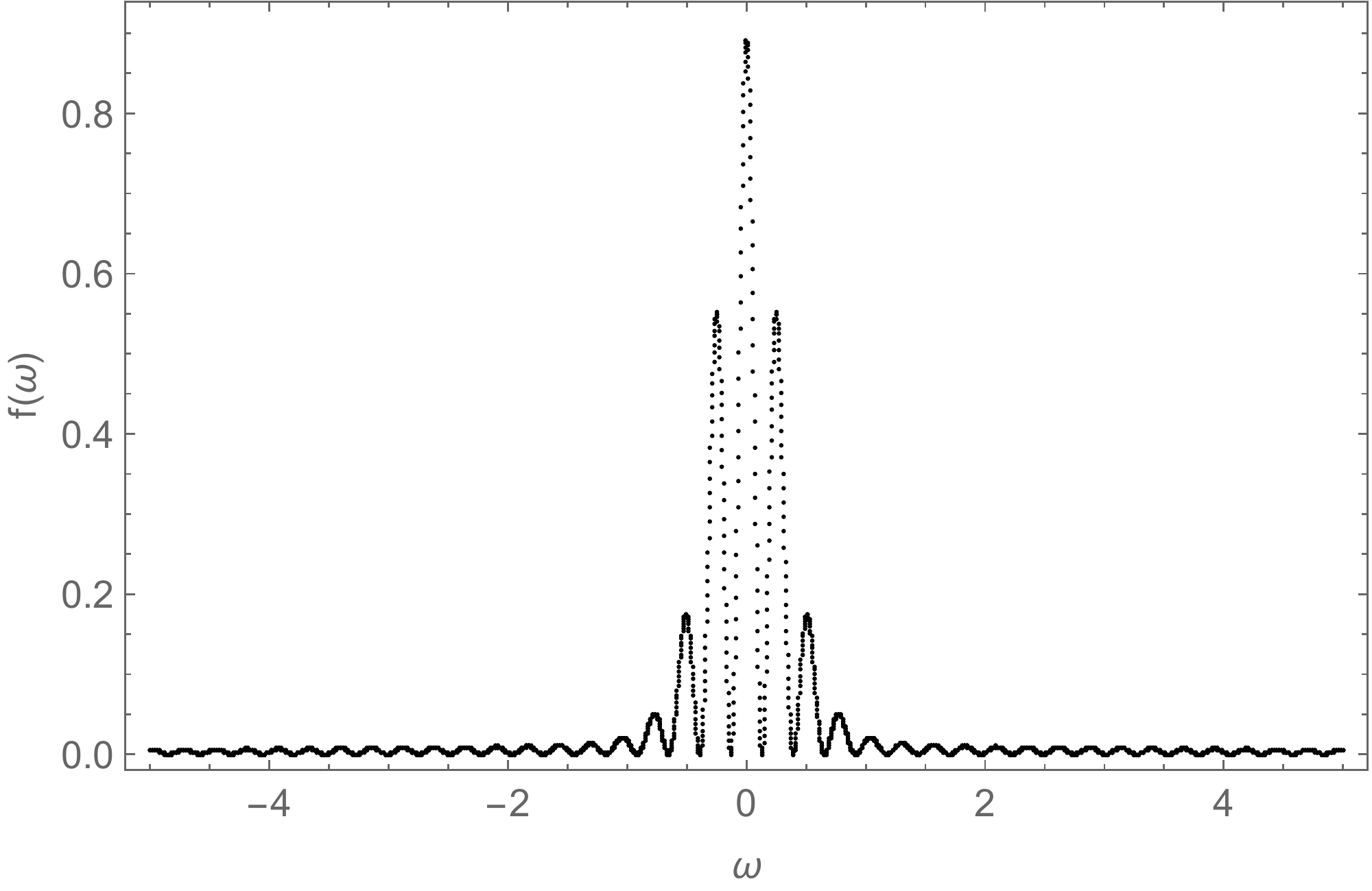}
\quad\quad
\includegraphics[width=2.99in]{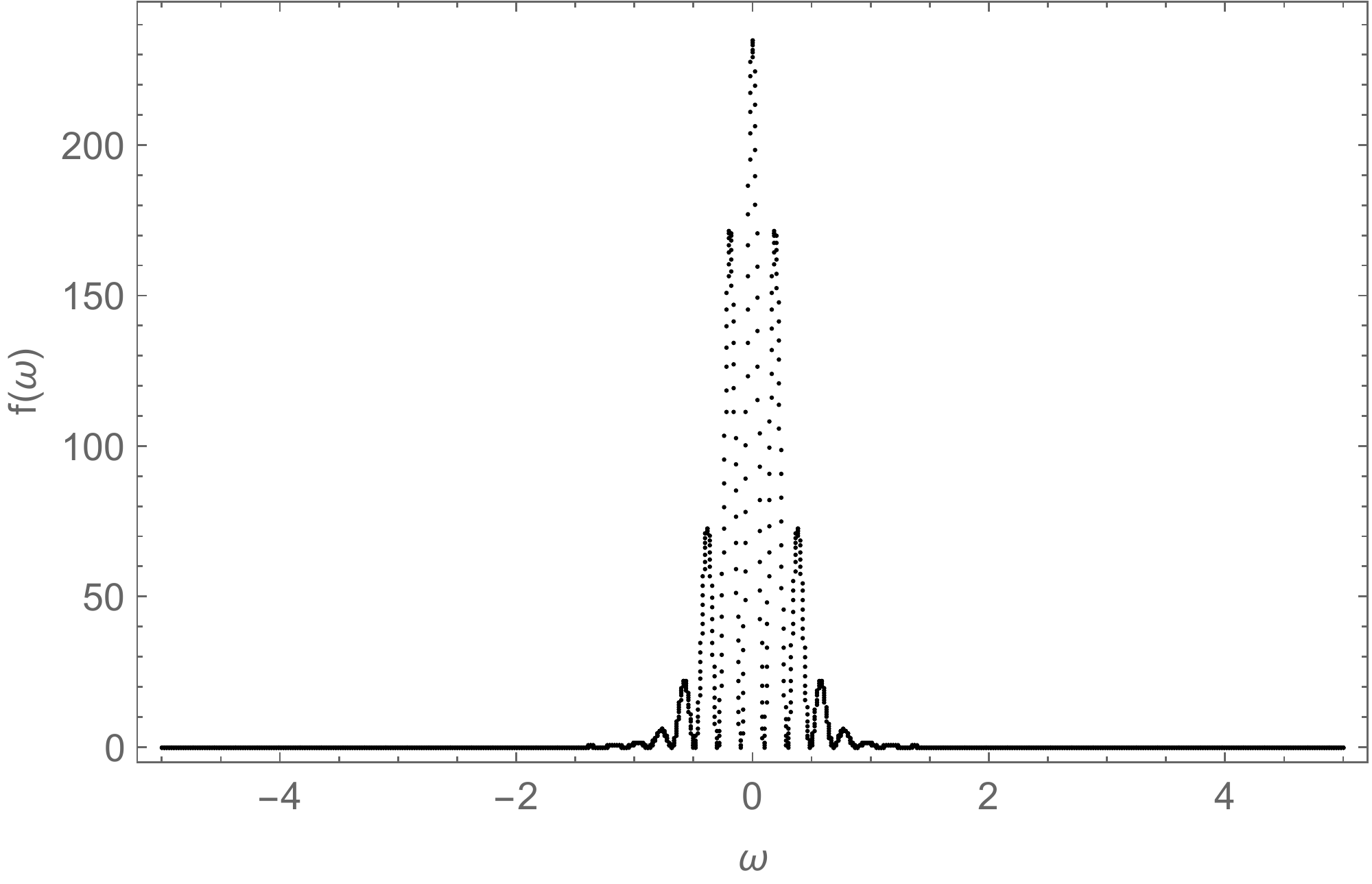}\\
\caption{\footnotesize Normalized structure factor ratio for the IR regime of the dynamical AdS/QCD holographic model, with warp factor compatible with linear Regge trajectories (\ref{cosmologica}), for $S=0$ (top left panel); $S=1$ (top right panel); $S=2$ (middle left panel); $S=3$ (middle right panel); $S=4$ (bottom left panel); $S=5$ (bottom right panel). }
\end{center}
\end{figure}
\begin{figure}[H]
\begin{center}
\includegraphics[width=2.99in]{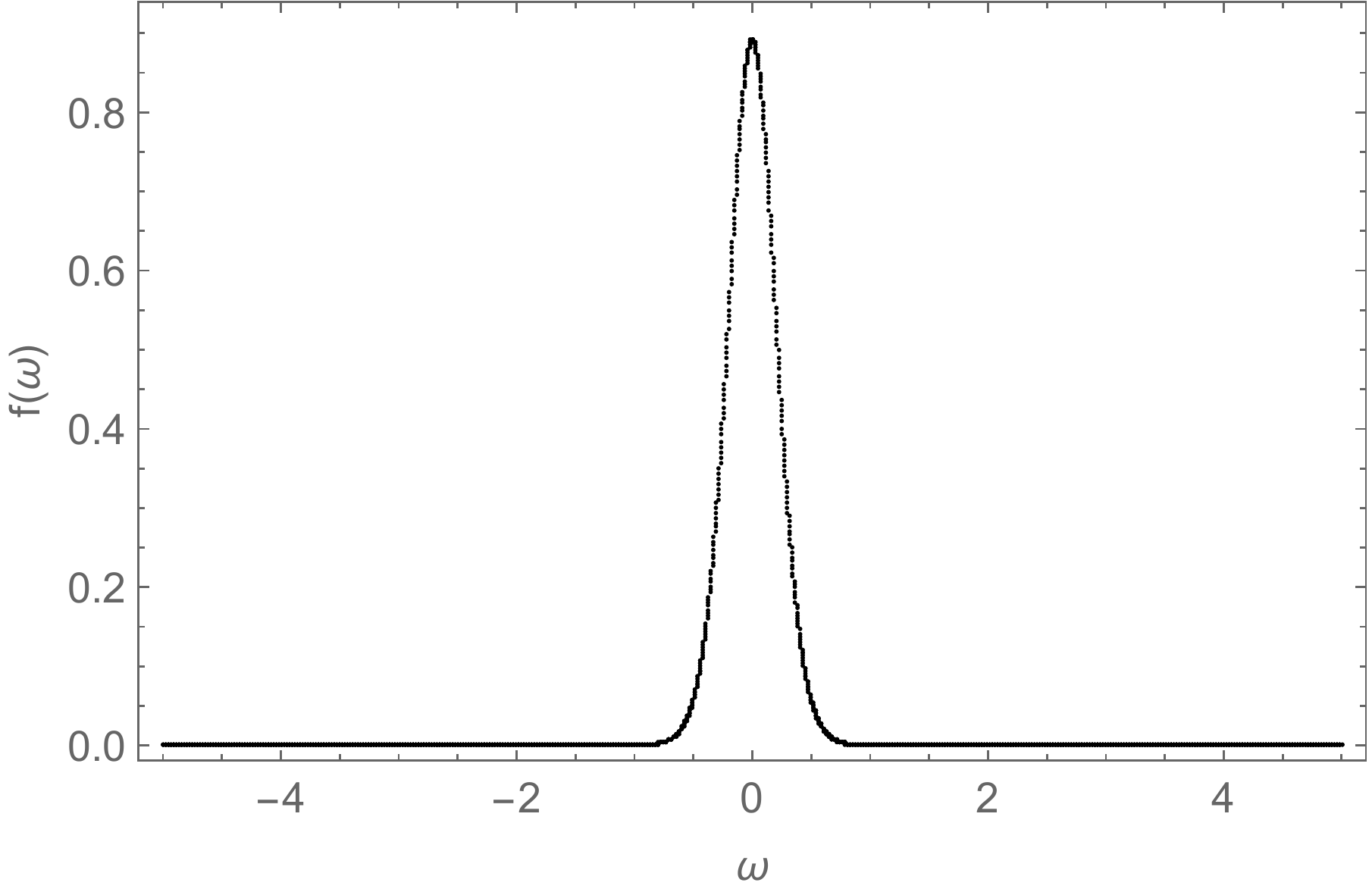}
\quad
\includegraphics[width=2.99in]{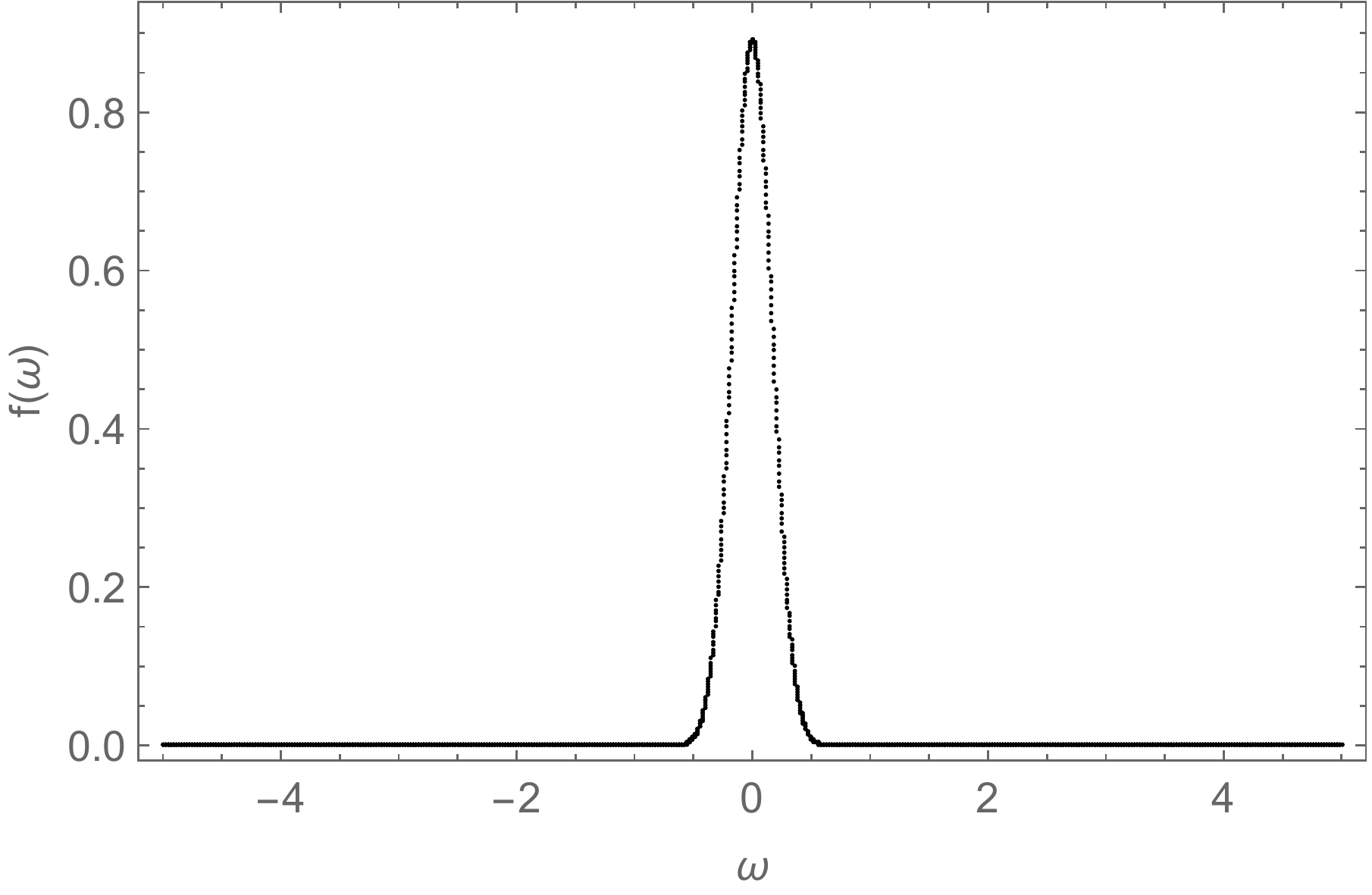}\\
\includegraphics[width=2.99in]{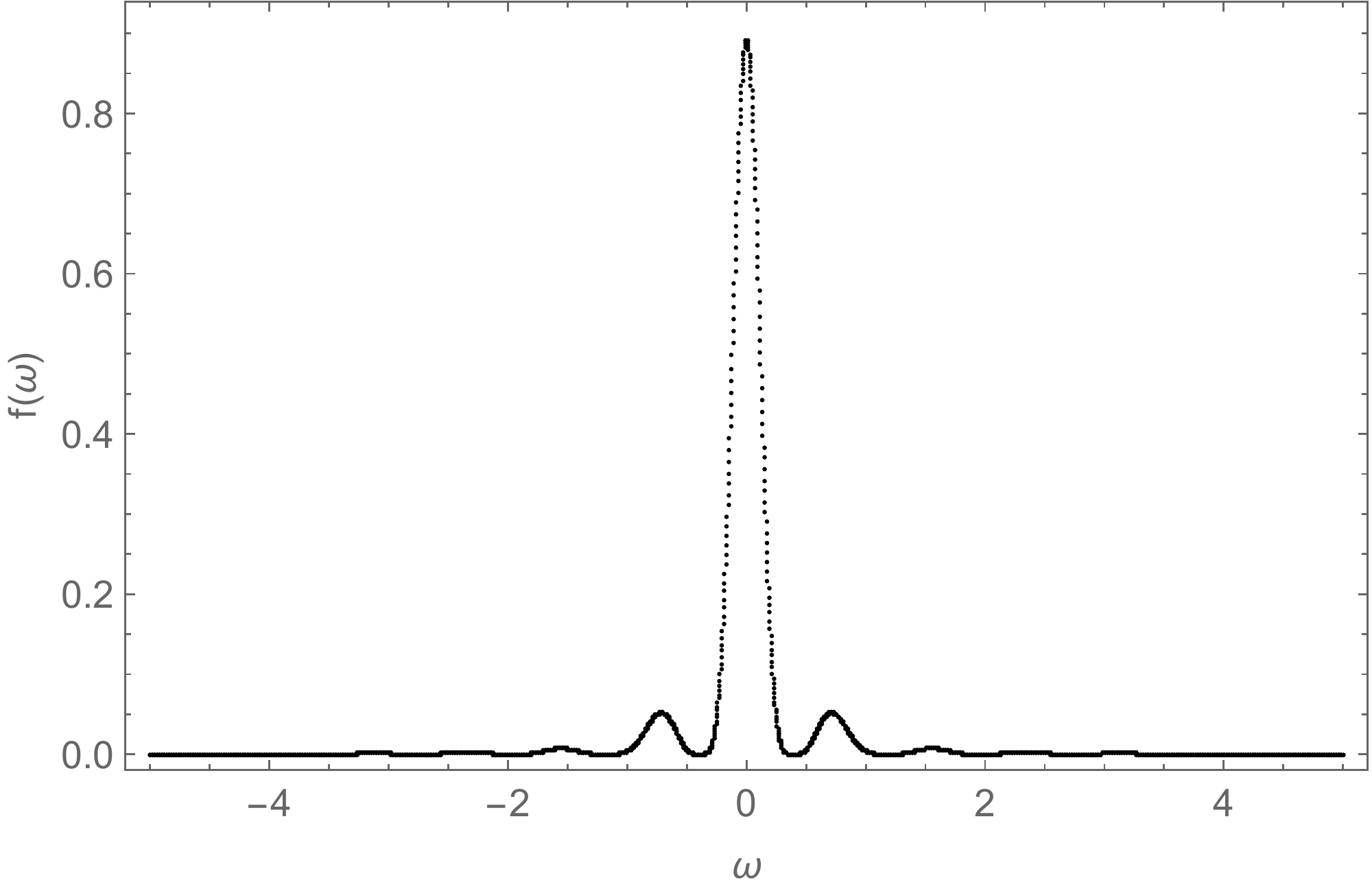}
\quad
\includegraphics[width=2.99in]{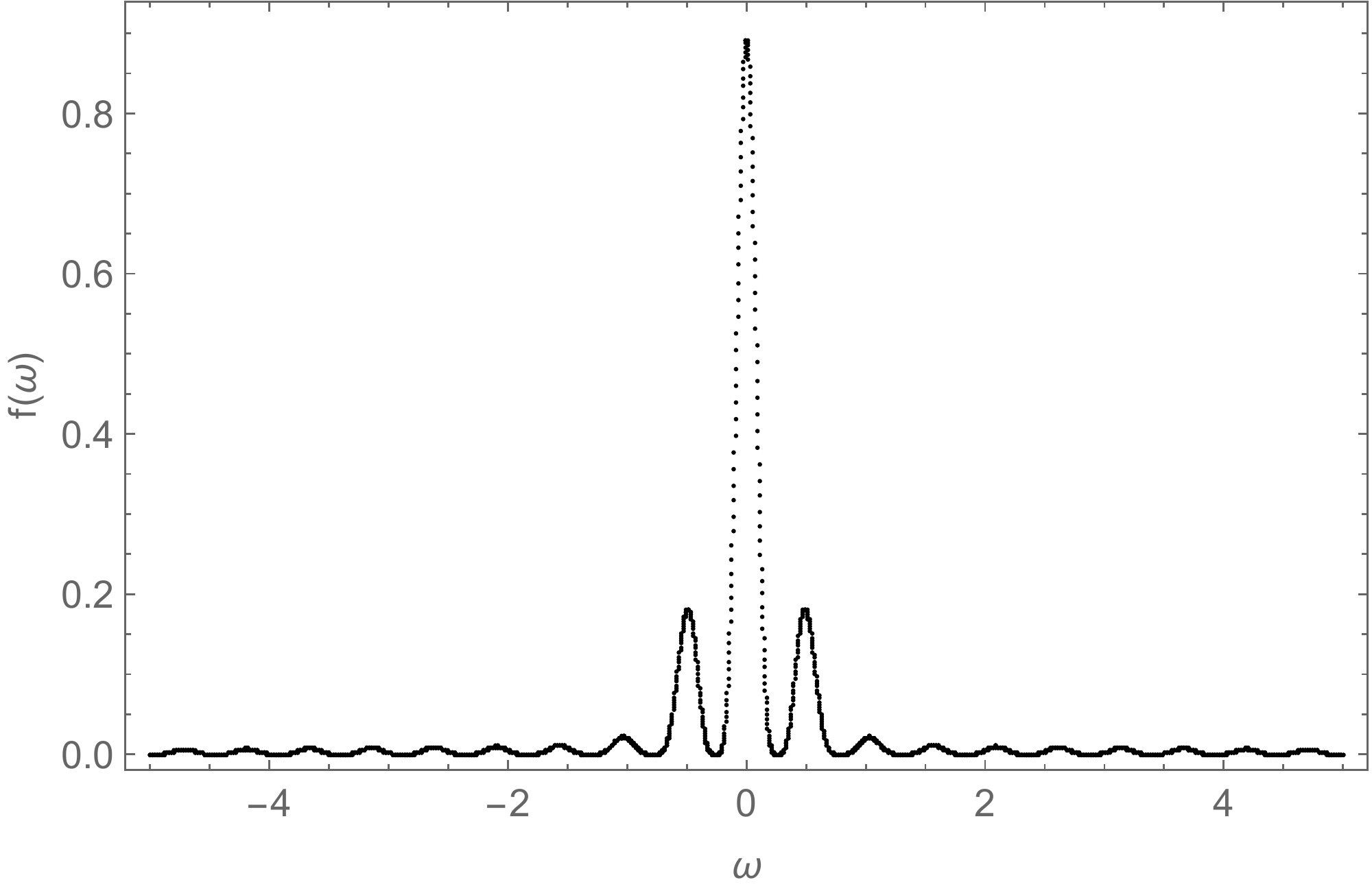}\\
\includegraphics[width=2.99in]{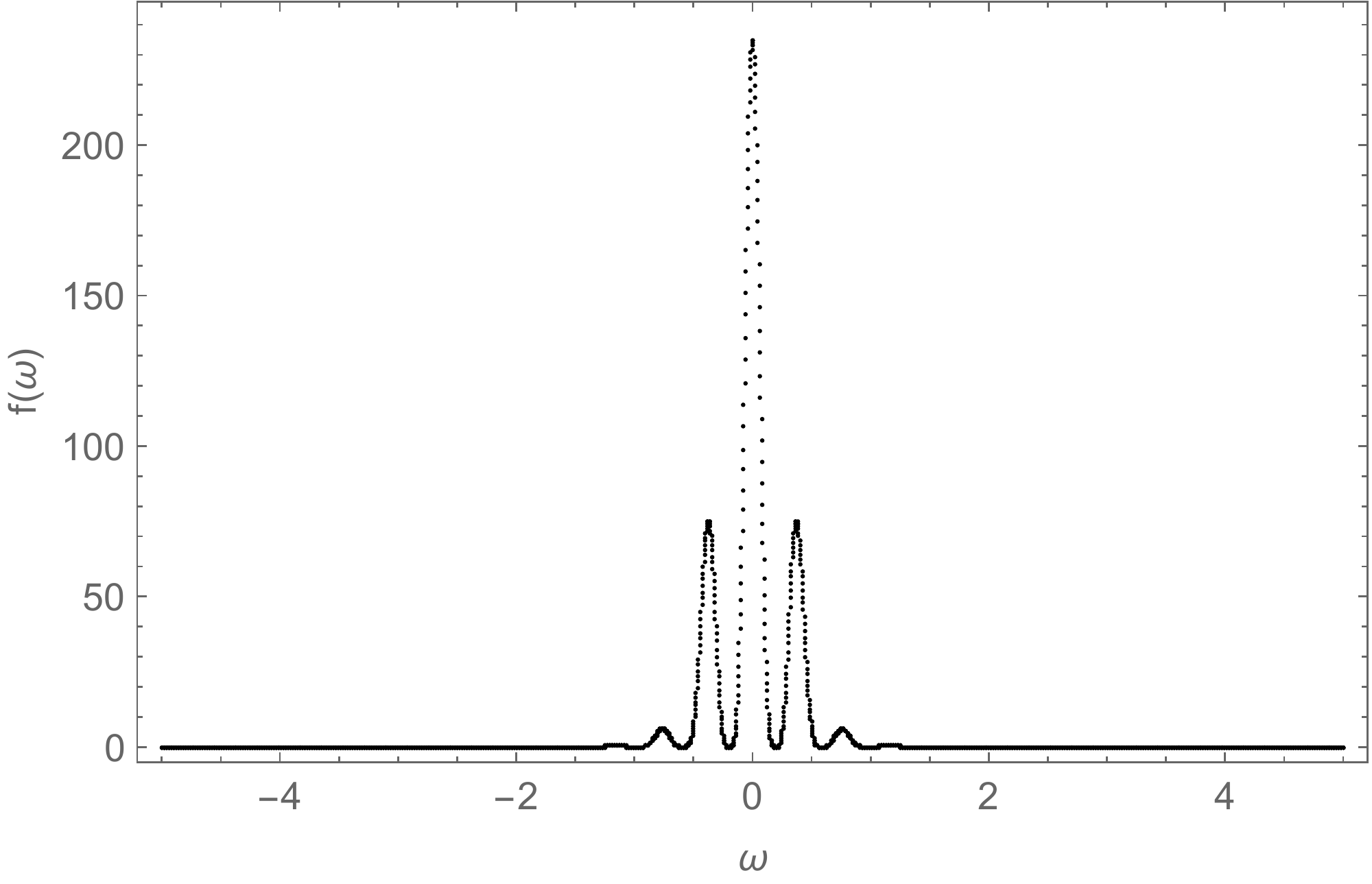}
\quad
\includegraphics[width=2.99in]{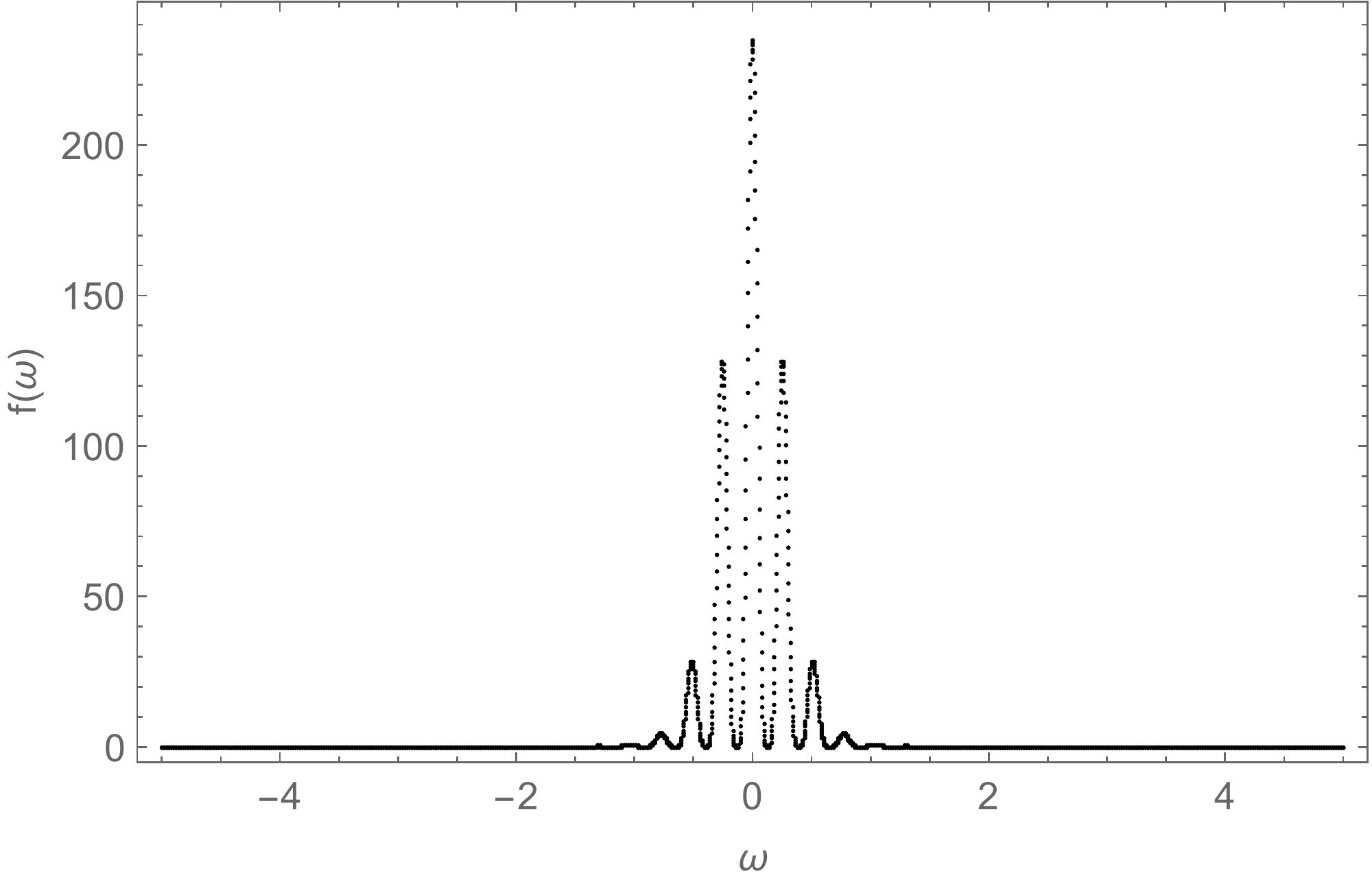}
\caption{\footnotesize  Normalized structure factor ratio for the UV limit of the dynamical AdS/QCD holographic model, with warp factor compatible with linear Regge trajectories (\ref{cosmologica}), for $S=0$ (top left panel); $S=1$ (top right panel); $S=2$ (middle left panel); $S=3$ (middle right panel); $S=4$ (bottom left panel); $S=5$ (bottom right panel).}
\end{center}
\end{figure}
\end{widetext}
 Thus, Eq. (\ref{collectivecoordinates}) can be used to
generate the normalized structure factor ratio in Eq. (\ref{collective1}), in order to obtain the entropic profile of the dynamical AdS/QCD holographic model. In fact, the conditional entropy is calculated for both the UV and IR limits and displayed in 
Figs. 3 and 4.
\begin{figure}[H]
\begin{center}
\includegraphics[width=2.99in]{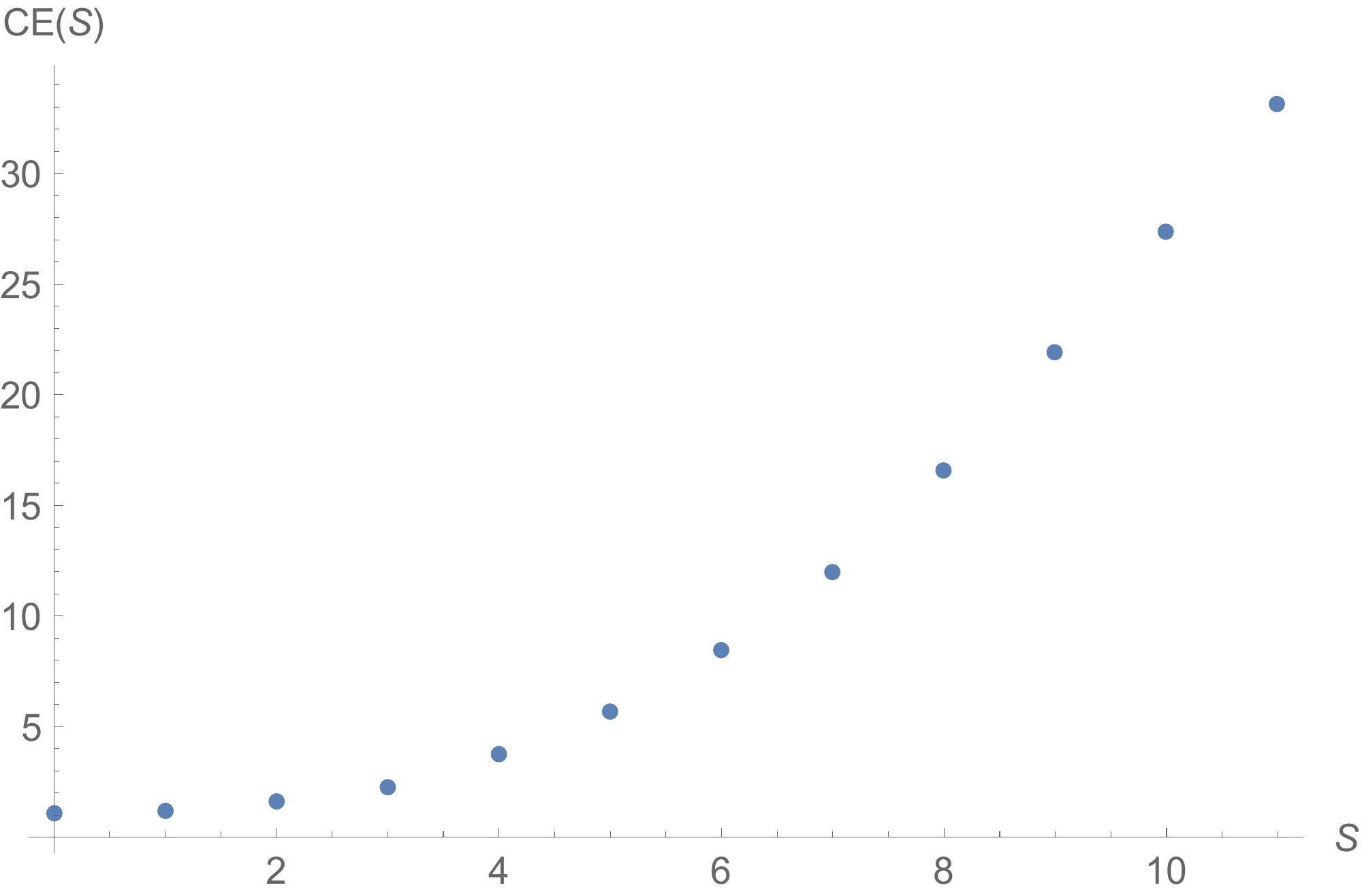}
\quad\quad
\caption{\footnotesize\; Conditional entropy (CE) as a function of the light-flavour meson spin, in the dynamical AdS/QCD holographic model, for the IR limit (here one adopted the notation CE($S$) as to not confuse the spin ($S$) with the CE ($S_c$)).}
\end{center}
\label{ce1}
\end{figure}

\begin{figure}[H]
\label{ce2}
\begin{center}
\includegraphics[width=2.99in]{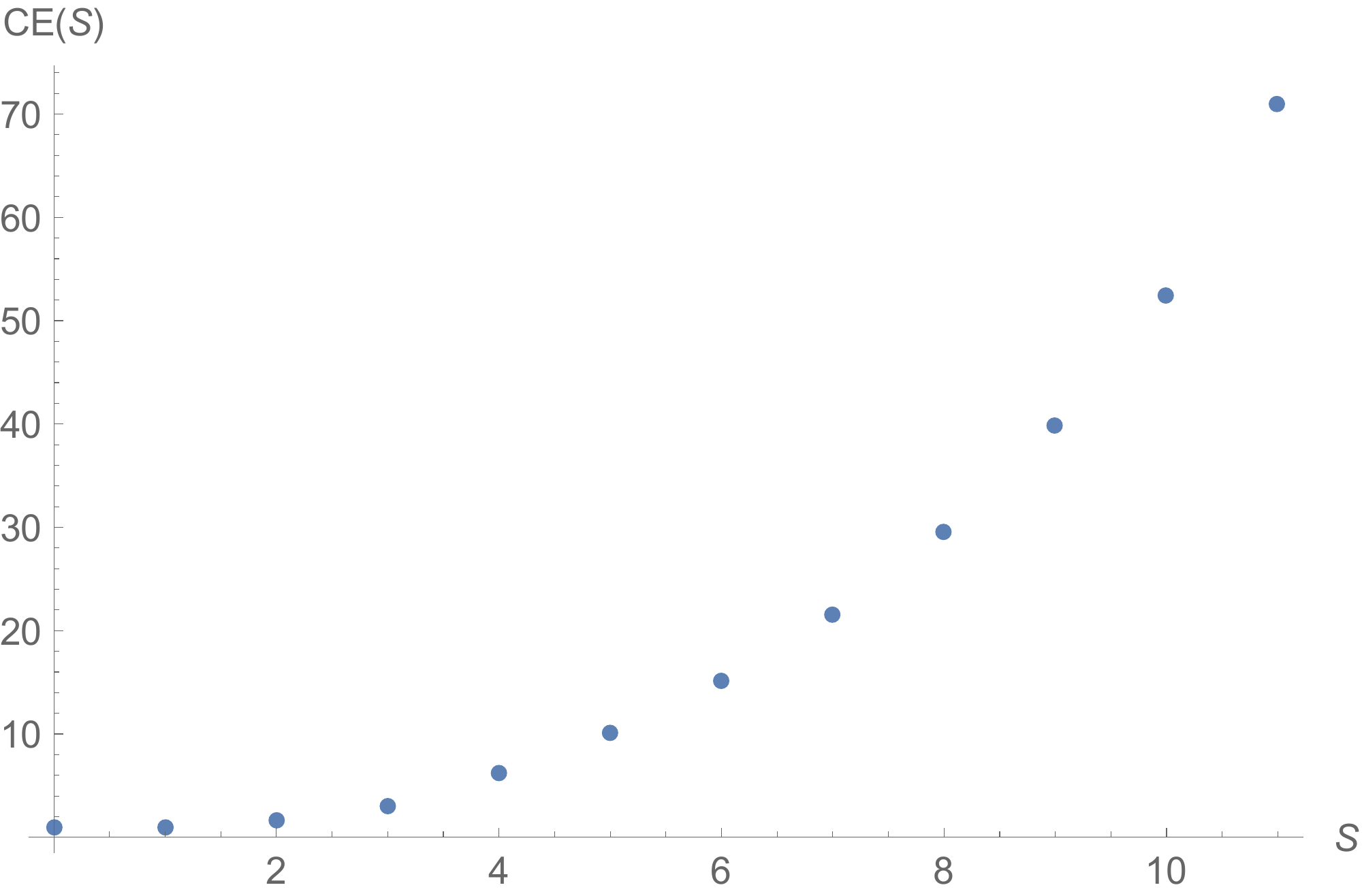}
\quad\quad
\caption{Conditional entropy (CE) as a function of the light flavour meson spin $S$, in the AdS/QCD holographic model, for the UV limit (here one adopted the notation CE($S$) as to not confuse the spin ($S$) with the CE ($S_c$)). }
\end{center}
\end{figure}
It is worth to emphasise that the CE for both UV and IR limits, depicted in Figs. 3 and 4, 
has a very close value for $S=0$ and $S=1$. Clearly, the higher is the spin, the higher is the CE, showing an increasing configurational instability as the light-flavour meson spin increases. Hence, the higher is the light-flavour mesons spins, the more unpredictable is the information content of the system. In this scenario, light-flavour mesons with lower spins are entropically more stable -- and then more frequent in Nature -- than higher spins mesonic states. It can explain the higher abundance of lower spin mesons in Nature, when compared to its higher spin counterparts \cite{PDG,pdg1}. 
One should indeed expect that the CE varies at different spins, since different
modes start to be active, for different spins. In particular, for lower spins, when the average field distribution is dominated 
by a few long-wavelength modes, one should expect the CE to be lower. 

 As final remarks, in Ref. \cite{SongHe2010}, the deformed $D_p-D_q$ soft-wall model was employed in the context of type II superstring theory. From the  asymptotically AdS$_5$ metric ($D_3$ system), and the Sakai-Sugimoto
model ($D_4-D_8$ system), the $D_p-D_q$ system was analysed and shown to not describe linear
trajectories of mesons. In this context, a  deformed $D_p-D_q$ soft-wall model in QCD was defined \cite{SongHe2010}: 
\begin{equation}
A(z)=-a_0 ~\mathrm{ln} z, \, ~\phi(z)=d_{0}\;\ln z+c_{2}z^{2}\,.\nonumber
\label{scenario1}
\end{equation}
The
effective 5D action for higher spin mesons described by tensor fields as the one before Eq. (\ref{BBB1}). Moreover, the defining solution
of the function $B=\left( 2S-1\right) A+\phi$, placed after Eq. (\ref{BBBB}), can take a slightly more general form 
\begin{eqnarray}
B=\phi-k (2\,S-1) A= \phi+c_0 (2\,S-1) \mathrm{ln}z.
\end{eqnarray}
The function  $B(z)$ similarly approaches logarithmically  asymptotic at the UV limit, and asymptotes $z^2$, in the IR regime.
The parameter $k$  depends upon the induced metric of the $D_q$ brane, reading $k$ reads \cite{SongHe2010}
$ 
k=-\frac{\left( p-3\right) \left( q-5\right) +4}{7-p},$ 
where $
c_0=k a_0= -\frac{\left( p-3\right) \left( q-5\right) +4}{2\left( p-5\right) 
}.$ 
The  parameters  of the $D_4-D_8$ system are $p=4$, $q=8$, implying that $k=7/3$, $a_0=3/2$, $c_0=7/2$, $d_0=-3/2$ and the curvature $\mathcal{R}\sim1/g_{\rm effective} = \frac{1}{6z^2\sqrt{\pi}}$, as in Table III of \cite{SongHe2010}. They are here adopted to plot the graphic:
\begin{figure}[H]
\begin{center}
\includegraphics[width=2.99in]{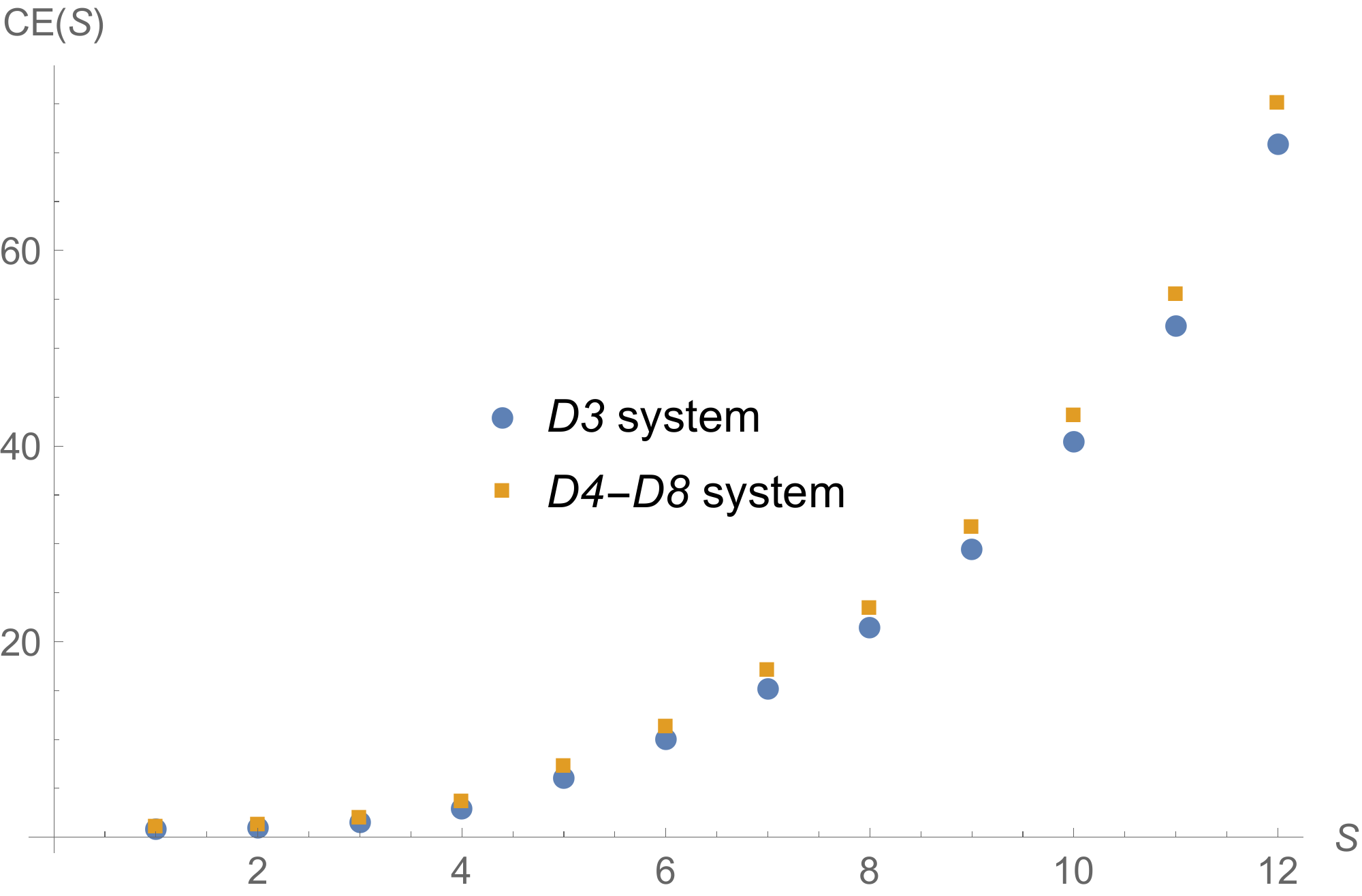}
\quad\quad
\caption{\footnotesize\; Conditional entropy (CE) as a function of the light-flavour meson spin, for the IR limit. The bullets depict Fig. \ref{ce1} for the $D_3$ system corresponding to the AdS metric, whereas the squares depict the  $D_4-D_8$ system, when  $p=4$, $q=8$, $k=7/3$, $a_0=3/2$, $c_0=7/2$, $d_0=-3/2$.}
\end{center}
\label{ce101}
\end{figure}
\begin{figure}[H]
\begin{center}
\includegraphics[width=2.99in]{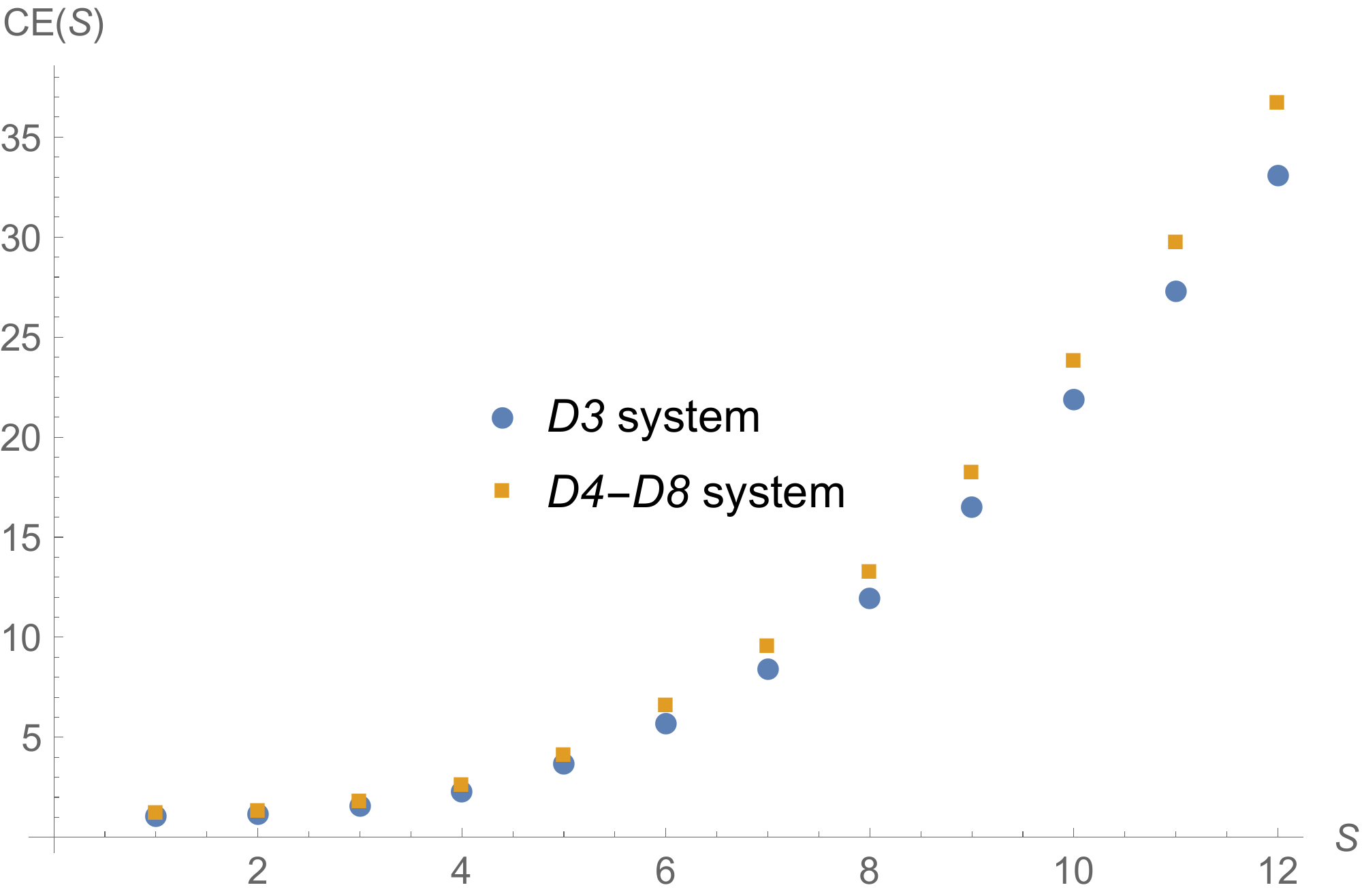}
\quad\quad
\caption{\footnotesize\; Conditional entropy (CE) as a function of the light-flavour meson spin, for the UV limit. The bullets depict Fig. \ref{ce2} for the $D_3$ system corresponding to the AdS metric, whereas the squares depict the  $D_4-D_8$ system, when  $p=4$, $q=8$, $k=7/3$, $a_0=3/2$, $c_0=7/2$, $d_0=-3/2$.}
\end{center}
\label{ce1001}
\end{figure}
It is worth to emphasize that Figs. 5 and 6 show that such kind of refinements 
do not alter the qualitative profile of Figs. 3 and 4. Still, it provides a quantitative tool for deriving the stability of mesonic states, that can be refined to encompass any kind of refinement. In fact, the $D_4-D_8$ system setup refines the results in Figs. 3 and 4 in around 6\%, for spin $S=12$ (in the IR limit), whereas around 7\% for spin $S=12$ (in the UV limit). Other models have similar corrections, however all models in Table III \cite{SongHe2010} can provide a maximum correction of $\sim 10\%$, irrespective of the meson spin, up to $S=15$.

\section{Concluding Remarks and Outlook}

The entropic information driven by the CE has been here studied in the dynamical AdS/QCD holographic model, realised by the solutions of the 5D Einstein-dilaton system that serves a dual background, for
holographic QCD. Our main result encodes a relationship between mesons with lower spins, compared with higher spin excited mesonic states, based upon the entropic information associated with the light-flavour mesonic system. This is corroborated by the analysis of the CE, calculated for the dynamical AdS/QCD holographic model, in the regime that is compatible with 
linear Regge trajectories, dictated by the warp factor (\ref{cosmologica}).
Our results, regarding the entropic information devised by the CE, are depicted in Figs. 3 and 4, respectively for the UV and IR limits. Furthermore, the normalized structure factor ratios in Figs. 1 and 2 evince interesting 
profiles, firstly noticed here, among the existing literature reporting the CE paradigm. Better ordered models, from the informational point of view, are related to the ones with lower spins, as shown in Figs. 3 and 4. In conformal coordinates, the field equations for the soft wall model are led to a 1-dimensional Schr\"odinger equation. Hence regarding the bulk extra dimension, the problem is reduced to the analysis of a potential in standard quantum mechanics, wherein the system is well known to tend to the ground state.

To determine the energy density in Eq. (\ref{eq.5}), the higher spin $S$ mesons contributions, due to large $N_c$ suppression, was not considered. When $S$ becomes very high, the energy density will get the comparable contribution from higher spin mesons. In fact, by following Subsection 24.3 of \cite{nastase}, we present the resulting computations with those corrections, encoding in the figure below the results. 
\begin{figure}[H]
\begin{center}
\includegraphics[width=2.99in]{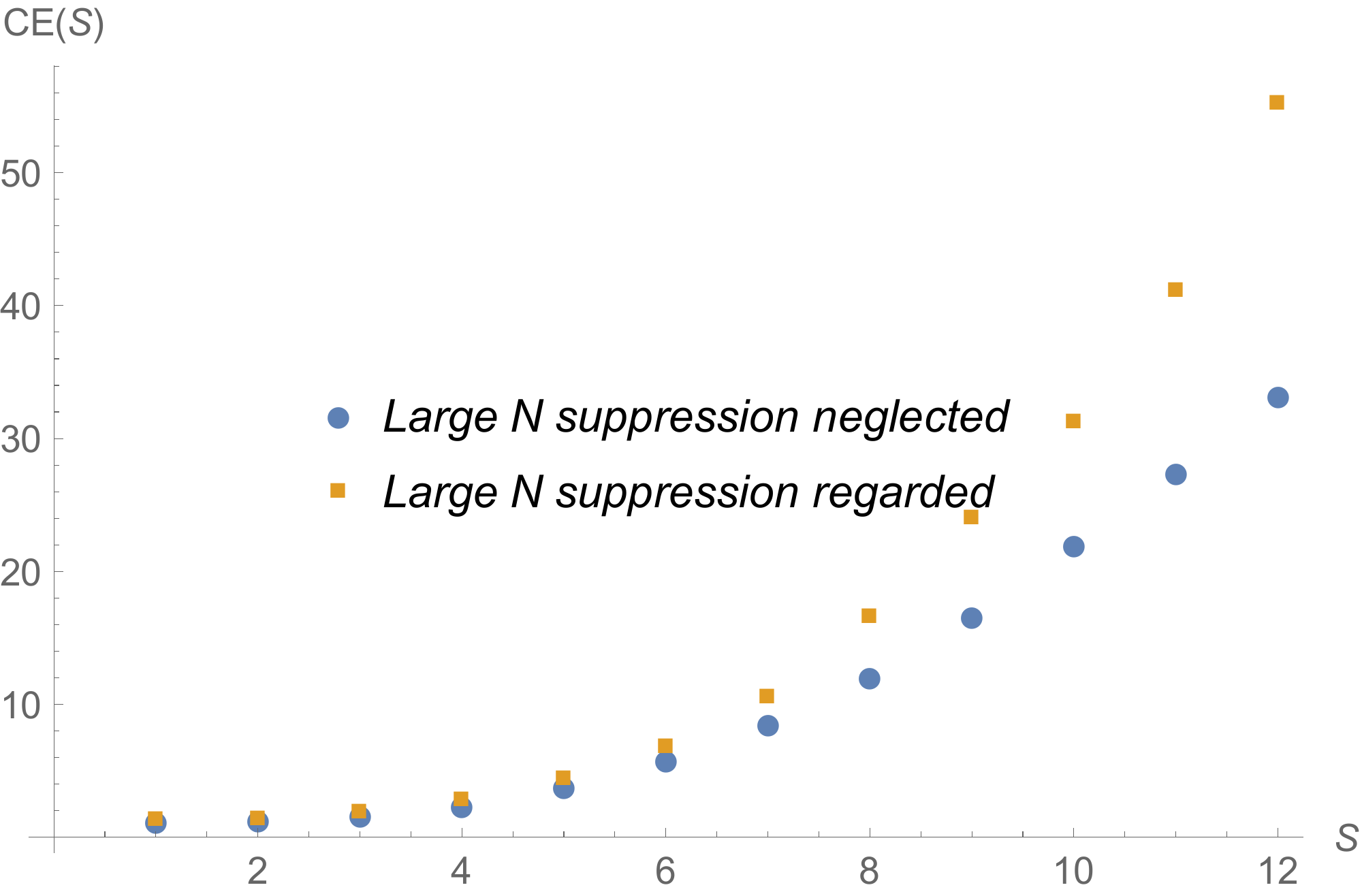}\label{ce333er}
\quad\quad
\caption{\footnotesize\; Conditional entropy (CE) as a function of the light-flavour meson spin, for the UV limit. The bullets depict Figure 3 in our paper and the squares represent the large $N_c$ suppression.}
\end{center}
\end{figure}
Since the corresponding beta function is defined perturbatively in the UV regime, Eq. (24.38) of \cite{nastase}, we illustrate this regime in Fig. 7.
In the model that regards Regge trajectories, corrected in the light of 
the QCD data \cite{Karch:2006pv,PDG}, it has been identified a positive correspondence 
the increasing of light-flavour meson spin values and increasing values the corresponding CE. Hence, each point of the graphics in Figs. 3 and 4 corresponds to a distinct mesonic particle spin. It implies that light-flavour mesons with very high spins, although not forbidden, by configurational entropic reasons, are very rare to be either produced or observed. In fact, in Table I, one notices the CE for some values of spins of light-flavour mesons:
\begin{center}
\begin{tabular}{||c|c||}
\hline\hline
\;Spin ($S$)\;& CE\\\hline\hline
10&\;\;$5.327\times 10^1$\\\hline
15&\;\;$8.43348\times 10^4$\\\hline
20&\; $6.26165\times10^6$\\\hline
25&\; $2.00055\times10^8$\\\hline
30&\;\;\;$2.28538\times10^{10}$\\
\hline\hline
\end{tabular}\\\medbreak\medbreak
Table I: CE for some values of spins of light-flavour mesons. \end{center} For instance, the CE for $S=30$ is $S_c \sim 2.28538 \times 10^{10}$, what makes a light-flavour meson of spin $S=30$ a configurationally unstable system with extremely high CE. This system has a huge entropic information content, what makes it to be practically impossible to be measured, detected or observed. If produced, the CE provides a quantitative apparatus to study the instability of high spin light-flavour mesons. 
Our results reinforce the outstanding usefulness of the CE framework, in particular, to 
analyse the configurational stability of physical systems. It consists into a pragmatic tool 
for experimental physicists to test the feasibility of potentially obtainable (QCD) data. 
This setup can be applied, moreover, to an extended range of particles. 

As the light flavor meson family has become increasingly abundant, the conditional entropy is closely related to the widths of observed high-spin states resonances, for spins $S=3,4,5,6$, as can be checked in Table I of Ref. \cite{Pang:2015eha}. 
Although more than twenty high-spin states are listed in \cite{pdg1}, their properties
are not currently well established. Therefore, the conditional entropy can work to determine how to categorize these high-spin states into meson families, shedding further light in the meson classification, since the experimental information of these high-spin states is not abundant.

\acknowledgments
The authors are grateful to Prof. Amilcar Queiroz, Prof. Nelson R. F. Braga and Prof. Henrique Boschi-Filho for fruitful discussions. 
The work of AEB is supported by the Brazilian Agencies FAPESP (grant 2015/05903-4) and CNPq (grant No. 300809/2013-1 and grant No. 440446/2014-7). RdR is grateful to CNPq (grants No. 303293/2015-2 and No. 473326/2013-2), and to FAPESP (grant 2015/10270-0) for partial financial support.

\end{document}